\documentclass{article}
\newif\ifexternal
\externaltrue

\PassOptionsToPackage{numbers}{natbib}
\usepackage{lmodern}
\usepackage[preprint]{neurips_2026}
\usepackage[english]{babel}

\usepackage{graphicx} 
\usepackage{hyperref}
\usepackage{amsmath}
\usepackage{amsfonts}
\usepackage{amssymb}
\usepackage{algorithm}
\usepackage{algorithmic}
\usepackage{comment}
\usepackage{multirow}
\usepackage{multicol}
\usepackage{tikz}
\usetikzlibrary{positioning, arrows.meta, shapes.geometric, fit, backgrounds}
\usepackage{ctable}
\usepackage{subcaption}
\usepackage{amsthm}
\theoremstyle{definition}

\usepackage{booktabs}
\usepackage{microtype}
\usepackage{mathtools}
\usepackage[utf8]{inputenc}
\usepackage[T1]{fontenc}
\usepackage{xcolor}

\title{Ready from Day 1: Population-Aware Coordination for Large-Scale Constrained Multi-Agent Systems}

\author{%
  Angel Wang \\
  Amazon \\
  \texttt{wangange@amazon.com}
  \And
  Dominique Perrault-Joncas \\
  Amazon \\
  \texttt{joncas@amazon.com}
  \And
  Alvaro Maggiar \\
  Amazon \\
  \texttt{maggiara@amazon.com}
  \And
  Dean Foster \\
  Amazon \\
  \texttt{foster@amazon.com} \\
  \And
  Carson Eisenach \\
  Amazon \\
  \texttt{ceisen@amazon.com}
}

\begin{document}

\maketitle

\begin{abstract}
In large-scale multi-agent systems with shared resource constraints, an upstream planner must iteratively evaluate candidate resource plans---assessing feasibility, aggregate response, and marginal cost---before committing to one. Lagrangian relaxation separates local decisions through a broadcast cost signal, but the planner still needs the cost-to-utilization response map to explore plan space, and this map depends on population composition that changes across planning cycles.
We propose \emph{population-aware coordination interfaces}: learned primal and dual maps, conditioned on compact population summaries, that the planner queries inside its iterative loop. The primal map predicts aggregate utilization under a proposed cost trajectory; the dual map predicts the cost trajectory for a target plan. By encoding response-relevant population structure, these maps remain reliable across evolving populations without per-cycle retraining, and support coordination of large populations from compact subsamples. We additionally cast Sim2Real transfer as a backtestable procedure, enabling evaluation before deployment. In a supply-chain capacity-control case study, population-aware interfaces reduce forecast error by 16--19\% and capacity violations by 20--51\% relative to population-unaware baselines under composition shift; 20K-agent cohorts support accurate coordination of 500K-agent populations; and simulator-trained primal maps achieve 11.1\% MAPE on real observations versus 13--24\% for baselines.
\end{abstract}

\section{Introduction}
Many modern systems consist of large populations of decentralized agents whose local actions collectively consume shared resources. Examples include distributed services sharing compute capacity, budget-constrained advertising systems, and supply-chain systems with storage and labor limits~\citep{cachon2003supply,federgruen1999coordination}. Even when local policies are effective in isolation, their aggregate behavior must be managed so that system-level resource budgets are respected. This creates a coordination problem for a {\it planner} -- solving problems such as resource pricing or allocation across agents -- who must choose feasible resource plans or coordination signals while reasoning about the aggregate outcomes they induce.

A centralized planner could in principle determine all individual actions, but this becomes difficult at large scale and may be incompatible with independently operated or already-trained policies. Dual-space coordination offers a more tractable alternative: by dualizing the shared constraint~\citep{boyd2011distributed,fisher1981lagrangean}, the planner converts the resource budget into a broadcast dual cost signal, which serves as the shadow price of the shared constraint. For any fixed cost trajectory, the cost-penalized objective separates across agents, so each agent responds locally while aggregate feasibility is governed through the shared dual variable. Unlike multi-agent policy-learning settings such as CTDE~\citep{lowe2017multi,oliehoek2016concise}, MADDPG~\citep{lowe2017multi}, and mean-field approaches~\citep{yang2018mean}, our goal is not to learn a joint policy, but to coordinate independently operated local policies through a shared signal.

This separability shifts the main difficulty to selecting the right resource plan and cost signal. Unlike execution-time cost control, which takes a resource plan as given and produces cost signals to enforce it, the upstream planner faces a broader problem: it does not know a good plan a priori, and must explore which plans are feasible, what aggregate responses they induce, and at what marginal cost before deciding whether to revise. A candidate plan may be infeasible at some horizon regardless of cost because past actions have already committed capacity, or feasible but with temporal spillovers that leave downstream periods over- or under-committed. The planner therefore needs repeated counterfactual queries of the cost-to-utilization response relationship as it compares and revises candidate plans.

This planning loop requires the response relationship in two directions. The forward (primal) map predicts aggregate utilization under a proposed cost trajectory, enabling feasibility evaluation and identification of temporal spillovers. The inverse (dual) map predicts the cost trajectory for a target plan, providing the marginal resource value the planner compares against its acquisition costs. Classical approaches, including iterative dual methods~\citep{boyd2004convex} and model predictive control~\citep{mayne2000constrained,garcia1989mpc,camacho2004model}, solve this through repeated inner-loop simulation of the full population, which is intractable when the loop must converge within operational time budgets. This motivates learned coordination interfaces: offline-trained models that support low-latency queries inside the planner's iterative loop~\citep{eisenach2024ncc}.

However, learned interfaces must remain reliable across the conditions under which the planner will query them. We focus on three challenges that existing work does not jointly address. First, aggregate response depends on population composition: two populations with similar aggregate statistics can respond differently to the same signal due to agent-level heterogeneity. Second, coordination must remain scalable as population size grows, since full-population evaluation can become impractical at scale. Third, coordination often begins before on-policy data is available -- each deployment is a cold start. Simulator-trained interfaces are useful only if the learned response map transfers to real population behavior~\citep{madeka2022deep,sinclair2023hindsight}, making it important to assess Sim2Real transfer before online use.

To address these challenges, we introduce \emph{population-aware coordination interfaces}: learned primal--dual models that condition both maps on compact, composition-aware summaries of the active population. This conditioning allows the planner to account for response-relevant heterogeneity across planning cycles without retraining the interface. We study this abstraction in supply-chain capacity control, where product-level replenishment policies respond to shared inbound-capacity costs while aggregate inbound flow must satisfy system-level limits.

Our key contributions are:

\textbf{(1) A unified primal--dual coordination abstraction.}
We formulate planner-facing coordination as paired primal and dual interfaces: a forward response map and an inverse cost-signal map that replace repeated simulation with low-latency queries inside the planner's iterative loop.

\textbf{(2) Population-aware coordination under population composition shift.}
Under controlled shifts in the distribution of agent responsiveness, population-aware interfaces
achieve 16--19\% lower forecast error and 20--51\% fewer capacity violations than
population-unaware baselines.

\textbf{(3) Scalable coordination via compact population representations.}
We show that accurate coordination does not require exhaustive per-agent querying of the full population. In our supply-chain setting, compact cohorts of roughly 20K agents support accurate prediction and control for target populations up to 500K agents.

\textbf{(4) Backtesting cold-start Sim2Real transfer.}
We treat Sim2Real transfer as a procedure that can be backtested: an aggregate forecaster trained in simulation is evaluated against real observations from past deployments. In a supply-chain case study, our proposed aggregate primal forecaster achieves 11.1\% MAPE on real observations, compared with 13--24\% for baselines, with a further 4--6\% gain from lightweight online calibration.

A complete review of related work is in Appendix~\ref{app:related_work}.

\section{Problem Formulation} \label{sec:formulation}

\subsection{Resource-Constrained Interactive Decision Process}

We consider a resource-constrained Interactive Decision Process (IDP)~\citep{madeka2022deep,eisenach2024ncc,sinclair2023hindsight,andaz2023learning} in which a population $S_t$ of agents acts at times $t=0,\ldots,T$.

\paragraph{State, Actions, and Utilization.}
The system state $x_t$ consists of local states $\{x_t^i\}_{i\in S_t}$, including endogenous variables that summarize the effects of past actions and coordination signals, and exogenous signals that evolve independently of the policy. Each agent $i \in S_t$ selects an action $a_t^i$ according to a local policy $\pi$. Under stochastic dynamics, resource utilization is generated through the system transition dynamics; we write $J_t^i = J^i(x_t^i,a_t^i) \in \mathbb{R}_+^m$, where $m$ is the number of shared resources. Aggregate utilization is:
\[
J_t = \sum_{i\in S_t} J_t^i \in \mathbb{R}_+^m.
\]

\paragraph{Constrained Coordination.}
At each planning cycle, an upstream planner specifies a capacity target $G_{0:T}=(G_0,\ldots,G_T)$, with $G_t\in\mathbb{R}_+^m$. A classical constrained-control formulation is
\begin{align}
\max_{\pi \in \Pi} \quad & \mathbb{E}\!\left[\sum_{t=0}^T \sum_{i \in S_t} U_t^i(x_t^i,a_t^i)\right] \notag \\
\text{s.t.} \quad & \mathbb{E}[J_t] \le G_t, \quad \forall\, t,
\end{align}
where $U_t^i(x_t^i,a_t^i)$ is the per-agent utility of action $a_t^i$ given state $x_t^i$, and inequalities are component-wise when $m>1$.

\paragraph{Dualization and Separability.}
Rather than solving the centralized problem directly in primal action space, we use dualization to obtain a broadcast cost signal. Introducing nonnegative dual variables $\lambda_t \in \mathbb{R}_+^m$ for the
capacity constraints $\mathbb{E}[J_t] \le G_t$ yields the Lagrangian
\[
\mathcal{L}(\pi,\lambda)
=
\mathbb{E}\!\left[
\sum_{t=0}^T \sum_{i\in S_t}
\left(
U_t^i(x_t^i,a_t^i)
-
\lambda_t^\top J_t^i
\right)
\right]
+
\sum_{t=0}^T
\lambda_t^\top G_t.
\]

For a given cost trajectory $\lambda_{0:T}$, the cost-penalized objective separates across agents: each agent internalizes the marginal resource penalty $\lambda_t^\top J_t^i$ and responds locally to the broadcast cost signal.

The planner's task is therefore to characterize the aggregate response induced by a broadcast trajectory:
\begin{equation}
\mathcal{J}_t(x_t,S_t,\lambda_t)
=
\sum_{i\in S_t}
\mathcal{J}_t^i(x_t^i,\lambda_t)
\in \mathbb{R}_+^m,
\label{eq:aggregate_response_map}
\end{equation}
where $\mathcal{J}_t^i(x_t^i,\lambda_t) \triangleq \mathbb{E}_{a_t^i\sim\pi(\cdot \mid x_t^i,\lambda_t)}[J^i(x_t^i,a_t^i)]$ is the expected utilization induced by agent $i$'s local policy response to the cost signal $\lambda_t$.

\paragraph{Supply-chain Application.}
In the supply chain capacity-control setting evaluated in this paper, agents correspond to product-level operational policies such as replenishment, sourcing, and liquidation decisions. The state $x_t^i$ encodes endogenous variables such as on-hand inventory and in-transit order quantities, and exogenous signals such as demand forecasts. Actions $a_t^i$ are local inventory procurement decisions that affect future consumption of shared operational resources such as inbound truck capacity and warehouse storage. Our experiments use a single shared resource dimension ($m=1$): inbound capacity for inventory arrivals. Here, $J_t^i$ is the inbound resource consumed by agent $i$ at time $t$, and $G_t$ is the per-period capacity limit.

\subsection{Coordination Interfaces and the Planning Loop}
We define a \emph{coordination interface} as the planner-facing abstraction for exploring candidate resource plans $G_{0:T}$. Over a planning horizon $L$, the interface exposes two maps:

\begin{equation}
J_{t:t+L} = \psi(x_t,S_t,\lambda_{t:t+L}), \qquad
\lambda_{t:t+L} = \phi(x_t,S_t,G_{t:t+L}).
\label{eq:coordination_interface}
\end{equation}

The primal map $\psi$ returns aggregate utilization induced by a proposed cost trajectory for the active population $S_t$. The dual map $\phi$ returns the cost trajectory associated with a target resource plan.

\paragraph{Planning Loop.}
Given the active population $S_t$ and an initial plan $G^{(0)}_{0:T}$, the planner can iteratively query the interface and revise the plan:
\begin{equation}
G^{(k+1)}_{0:T} = \mathcal{P}(G^{(k)}_{0:T},\; J^{(k)}_{t:t+L},\; \lambda^{(k)}_{t:t+L}),
\label{eq:planning_loop}
\end{equation}
where $\mathcal{P}$ is the planner's revision rule, incorporating its own cost structure (e.g., marginal resource-acquisition cost) and operational constraints~\citep{maggiar2024cpp}. The planner uses $J^{(k)}$ to assess feasibility and identify temporal spillovers, and $\lambda^{(k)}$ to evaluate whether the implied resource price justifies the plan. Because this loop must converge within operational time budgets and runs periodically over evolving populations, the interface must be both fast and robust to population-composition changes.

\paragraph{Learned-Map Implementation.}
We implement the interface with learned maps $\psi_\theta$ and $\phi_\theta$, trained offline from simulator data to predict $\hat{J}_{t:t+L}$ and $\hat{\lambda}_{t:t+L}$, respectively. The goal is reliable counterfactual response across plans and evolving populations; the planner $\mathcal{P}$ is treated as given.

\subsection{Training Data from a Simulator}

We use a simulation environment as a counterfactual data generator, following the Exogenous-IDP framework of~\citep{madeka2022deep,eisenach2024ncc}. Exogenous processes can be replayed or sampled independently of the policy, while endogenous state variables are simulated forward because they depend on past actions. This separation allows the simulator to evaluate alternative coordination signals under controlled conditions, generating the diverse plan--response pairs needed to train the interfaces.

The simulator produces rollout data
\[
(x_{0:T},\; S_{0:T},\; G_{0:T},\; \lambda_{0:T},\; J_{0:T}),
\]
where each rollout specifies a target population, a target resource plan, and a coordination signal. The simulator applies $\lambda_{0:T}$ to the local policies, simulates the resulting actions and endogenous state transitions, and records aggregate utilization $J_{0:T}$.

The primal interface is trained offline as a supervised forward-response model: given $\lambda_{t:t+L}$, it predicts aggregate utilization $\hat{J}_{t:t+L}$ and minimizes prediction error against the simulated response $J_{t:t+L}$. The dual interface is trained via a closed-loop direct-backprop procedure~\citep{madeka2022deep,eisenach2024ncc}: given a target resource plan $G_{t:t+L}$, it predicts $\lambda_{t:t+L}$, which enters the fixed local policies as a resource-use penalty in the differentiable simulator, and the induced utilization is compared with the target using a capacity-violation loss. Gradients of this loss flow back through the simulator to update the dual model. Formal IDP formulation and simulator construction details are provided in Appendix~\ref{app:exo_idp}; full training procedures are provided in Appendix~\ref{app:training}.

\section{Population-Aware Coordination Interfaces} \label{sec:model_class}
\subsection{Population-Dependent Response}
\label{subsec:population_response}
Aggregate utilization is a sum of per-agent responses induced by the local policy $a_t^i\sim\pi(\cdot \mid x_t^i,\lambda_t)$. An agent's responsiveness to the broadcast signal is captured by its cost sensitivity $\nabla_{\lambda_t}\mathcal{J}_t^i(x_t^i,\lambda_t)$. At the population level,
\begin{equation}
\nabla_{\lambda_t}\mathcal{J}_t(x_t,S_t,\lambda_t)
=
\sum_{i\in S_t}
\nabla_{\lambda_t}\mathcal{J}_t^i(x_t^i,\lambda_t).
\label{eq:population_sensitivity}
\end{equation}
Thus, the response map depends on the mixture of agent-level sensitivities in $S_t$: changing this mixture changes the aggregate response to the same broadcast signal. We refer to this as \emph{population shift}. In the planning context, population shift arises both from natural evolution of the active population between planning cycles and from the planner's choice to coordinate different sub-populations within a cycle.

It is infeasible to learn a separate response model for every possible active population, since $S_t$ can vary combinatorially across planning cycles. Instead, we learn a shared interface that encodes $S_t$ through a compact population summary, enabling generalization across composition changes without retraining.

\subsection{Interface Architecture} \label{subsec:architecture}
We implement all coordination interfaces with the same base encoder--decoder architecture; variants differ only in how they represent the active population $S_t$.

\subsubsection*{Baselines}
The \textbf{Global Aggregate model} represents the active population $S_t$ using population-level summary features and predicts aggregate utilization directly. It is efficient but population-unaware: it cannot distinguish populations with similar aggregate statistics but different composition.

The agent-level \textbf{Bottom-Up model} represents $S_t$ through independent per-agent predictions that are aggregated by summation. This preserves agent-level heterogeneity but scales linearly with population size and cannot share information across agents when estimating aggregate response.

\subsubsection*{Population-Aware Interfaces}
Unlike the baselines, population-aware variants implement the coordination interfaces through a compact population summary.

Each agent is first mapped to an embedding
\begin{equation}
e_t^i = \mathrm{Enc}_{\theta}(x^i_{t-H:t}),
\label{eq:agent_embedding}
\end{equation}
where the encoder summarizes the agent's recent local history over the past $H$ steps. The agent embeddings are then pooled into a compact population representation
\begin{equation}
z_t = \rho_{\theta}\!\left(\{e_t^i\}_{i\in S_t}\right),
\label{eq:population_embedding}
\end{equation}
which captures the behavioral composition of the active population.

The decoder combines $z_t$ with the planner-provided trajectory to produce the corresponding coordination-interface output:
\[
\hat{J}_{t:t+L}
=
\mathrm{Dec}^{\mathrm{primal}}_\theta(z_t,\lambda_{t:t+L}),
\qquad
\hat{\lambda}_{t:t+L}
=
\mathrm{Dec}^{\mathrm{dual}}_\theta(z_t,G_{t:t+L}).
\]

The population embedding \(z_t\) conditions the decoder on the response-relevant composition of \(S_t\). For the primal head, the decoder combines \(z_t\) with the proposed cost trajectory to predict normalized utilization, which is rescaled by a population-scale factor to yield absolute aggregate utilization. For the dual head, the decoder combines \(z_t\) with the target capacity plan to predict a broadcast cost trajectory that is applied directly. Architecture details are provided in Appendix~\ref{app:architecture}.

Figure~\ref{fig:architecture_comparison} illustrates the architecture and two variants for constructing the population summary $\rho_\theta$. The \textbf{Population-Embedding (per-Agent) Aggregate} variant applies attention directly over agent embeddings, allowing the model to infer composition structure from data (Fig.~\ref{fig:architecture_comparison}, (a)). The \textbf{Population-Embedding (Bucketized) Aggregate} variant first groups agents into response-relevant segments, summarizes each bucket separately, and includes prevalence weights that encode the population mixture (Fig.~\ref{fig:architecture_comparison}, (b)). We use the structured bucketized variant as the primary model in Section~\ref{sec:experiments}, because its explicit composition representation improves generalization under population-composition shift; ablations are provided in Appendix~\ref{app:ablations}.

\begin{figure}[htbp]
\centering
\resizebox{\textwidth}{!}{%
\begin{tikzpicture}[
    node distance=0.5cm and 0.6cm,
    box/.style={rectangle, draw, minimum height=0.6cm, minimum width=1.3cm, font=\small},
    embbox/.style={rectangle, draw, fill=blue!15, minimum height=0.5cm, minimum width=0.9cm, font=\tiny, align=center},
    attention/.style={rectangle, draw, fill=orange!30, minimum width=0.6cm, minimum height=0.7cm, font=\tiny, align=center, rotate=-90},
    decoder/.style={rectangle, draw, fill=green!20, minimum width=0.8cm, minimum height=0.7cm, font=\tiny, align=center, rotate=-90},
    bucket/.style={rectangle, draw, dashed, rounded corners, fill=gray!10},
    costbox/.style={rectangle, draw, fill=purple!20, minimum height=0.5cm, minimum width=1.2cm, font=\tiny, align=center},
    planbox/.style={rectangle, draw, fill=purple!20, minimum height=0.5cm, minimum width=1.2cm, font=\tiny, align=center},
    arrow/.style={->, >=stealth, thick},
]

\node[font=\small\bfseries] at (-3.5, 4.2) {(a) Population-Embedding (per-Agent)};

\node[embbox] (e1L) at (-5, 2)   {$e^1_t\!=\!f(x^1_t)$};
\node[embbox] (e2L) at (-5, 1.3) {$e^2_t\!=\!f(x^2_t)$};
\node at (-5, 0.8) {\scriptsize$\vdots$};
\node[embbox] (enL) at (-5, 0.2) {$e^n_t\!=\!f(x^n_t)$};

\node[attention] (attnL) at (-3.3, 1.15) {Attention};
\coordinate (forkL) at (-2.7, 1.15);

\node[decoder]  (decL)  at (-2,   2.35) {Dec$^{\mathrm{P}}$};
\node[costbox]  (costL) at (-2,   3.45) {Cost $\lambda_{t:t+L}$};
\node[box, fill=yellow!30] (JL) at (-0.6, 2.35) {$\hat{J}_{t:t+L}$};

\node[decoder]  (decDualL) at (-2,   -0.05) {Dec$^{\mathrm{D}}$};
\node[planbox]  (planL)    at (-2,   -1.15) {Plan $G_{t:t+L}$};
\node[box, fill=yellow!30] (lambdaL) at (-0.6, -0.05) {$\hat{\lambda}_{t:t+L}$};

\draw[arrow] (e1L) -- (attnL);
\draw[arrow] (e2L) -- (attnL);
\draw[arrow] (enL) -- (attnL);
\draw[arrow] (attnL) -- (forkL);
\draw[arrow] (forkL) -- (-2.7,  2.35) -- (decL);
\draw[arrow] (forkL) -- (-2.7, -0.05) -- (decDualL);
\draw[arrow] (costL)    -- (decL);
\draw[arrow] (decL)     -- (JL);
\draw[arrow] (planL)    -- (decDualL);
\draw[arrow] (decDualL) -- (lambdaL);

\node[font=\small\bfseries] at (5.5, 4.2) {(b) Population-Embedding (Bucketized)};





\node[bucket, minimum width=3.7cm, minimum height=1.3cm] (bucket1) at (3.0, 2.1) {};
\node[font=\tiny, above] at (3.0, 2.75) {Bucket 1};

\node[bucket, minimum width=3.7cm, minimum height=1.3cm] (bucketK) at (3.0, -0.15) {};
\node[font=\tiny, above] at (3.0, 0.50) {Bucket $K$};

\node[embbox] (e1R)  at (2.2,  2.4) {$e^1_t\!=\!f(x^1_t)$};
\node[embbox] (e2R)  at (2.2,  1.8) {$e^2_t\!=\!f(x^2_t)$};
\node[embbox] (en1R) at (2.2,  0.15) {$e^{n\!-\!1}_t\!=\!f(x^{n-1}_t)$};
\node[embbox] (enR)  at (2.2, -0.45) {$e^n_t\!=\!f(x^n_t)$};

\node[attention] (attn1R)  at (4.1, 2.1) {Attn};
\node[attention] (attnKR)  at (4.1, -0.15) {Attn};

\node at (3.0, 1.05) {\small$\vdots$};

\node[attention] (crossAttnR) at (5.6, 1.05) {Attention};
\coordinate (forkR) at (6.2, 1.05);

\node[decoder]  (decR)  at (6.9,  2.25) {Dec$^{\mathrm{P}}$};
\node[costbox]  (costR) at (6.9,  3.35) {Cost $\lambda_{t:t+L}$};
\node[box, fill=yellow!30] (JR) at (8.2, 2.25) {$\hat{J}_{t:t+L}$};

\node[decoder]  (decDualR) at (6.9,  -0.15) {Dec$^{\mathrm{D}}$};
\node[planbox]  (planR)    at (6.9,  -1.25) {Plan $G_{t:t+L}$};
\node[box, fill=yellow!30] (lambdaR) at (8.2, -0.15) {$\hat{\lambda}_{t:t+L}$};

\draw[arrow] (e1R)  -- (attn1R);
\draw[arrow] (e2R)  -- (attn1R);
\draw[arrow] (en1R) -- (attnKR);
\draw[arrow] (enR)  -- (attnKR);
\draw[arrow] (attn1R) -- (crossAttnR);
\draw[arrow] (attnKR) -- (crossAttnR);
\draw[arrow] (crossAttnR) -- (forkR);
\draw[arrow] (forkR)      -- (6.2,  2.25) -- (decR);
\draw[arrow] (forkR)      -- (6.2, -0.15) -- (decDualR);
\draw[arrow] (costR)    -- (decR);
\draw[arrow] (decR)     -- (JR);
\draw[arrow] (planR)    -- (decDualR);
\draw[arrow] (decDualR) -- (lambdaR);

\end{tikzpicture}%
}
\caption{\small Population-aware forecaster architectures.
(a)~Population-Embedding (per-Agent) Aggregate: agent embeddings $e^i_t\!=\!f(x^i_t)$ are pooled via attention and then decoded.
(b)~Population-Embedding (Bucketized) Aggregate: within-bucket attention is followed by cross-bucket attention before decoding.
The population summary is passed to the selected decoder head: $\mathrm{Dec}^{\mathrm{P}}$ (primal) or $\mathrm{Dec}^{\mathrm{D}}$ (dual).}
\label{fig:architecture_comparison}
\end{figure}

\subsection{Generalization under Composition Shift}
\label{subsec:dist_shift}

Population-aware architectures address the representation problem: they provide a compact summary through which the interface can condition on $S_t$. Generalization under composition shift also requires data coverage: training rollouts must vary the population mixture enough for the model to learn how aggregate response changes with composition. We therefore construct shifted training populations by reweighting agents along response-relevant attributes.

\begin{figure}[htbp]
\centering
\begin{subfigure}{0.48\textwidth}
    \centering
    \includegraphics[width=\textwidth]{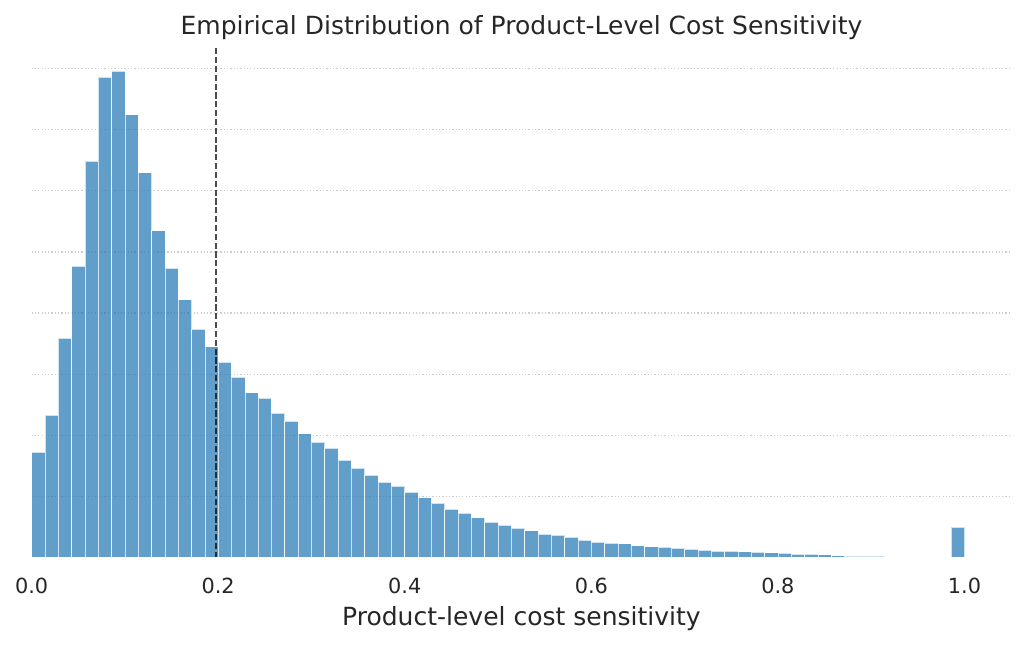}
    \caption{Cost sensitivity distribution}
    \label{fig:sensitivity_hist_main}
\end{subfigure}
\hfill
\begin{subfigure}{0.48\textwidth}
    \centering
    \includegraphics[width=\textwidth]{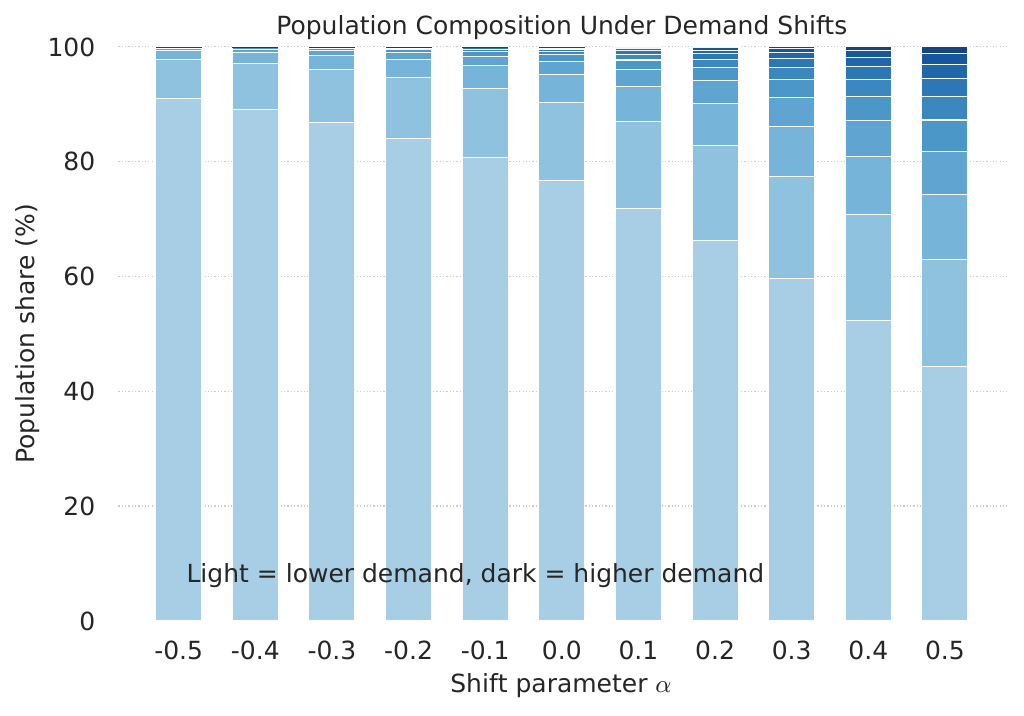}
    \caption{Population composition under $\alpha$-shifts}
    \label{fig:alpha_shift_main}
\end{subfigure}
\caption{\small (a) Distribution of agent-level cost sensitivity across 500K agents, showing a right-skewed tail of highly responsive agents. (b) Population composition under $\alpha$-shifted demand-decile mixtures in the supply chain setting: positive $\alpha$ upweights high-demand products, while negative $\alpha$ upweights low-demand products.}
\label{fig:ood_drivers}
\end{figure}

Agents are partitioned into quantile buckets along a selected attribute, and a shift parameter $\alpha$ moves population mass toward higher- or lower-value segments (see Appendix~\ref{app:population_shift} for the reweighting details). In the supply-chain setting, agent-level cost sensitivity is highly heterogeneous (Fig.~\ref{fig:ood_drivers}a); demand and unit economics are primary response-relevant attributes. We use demand-decile shifts as the main construction in the experiments: positive $\alpha$ upweights high-demand products, while negative $\alpha$ upweights low-demand products (Fig.~\ref{fig:ood_drivers}b).

We also generate target plans $\{G^{(n)}_{0:T}\}_{n=1}^N$ and associated cost trajectories $\{\lambda^{(n)}_{0:T}\}_{n=1}^N$ using the procedure described in Appendix~\ref{app:training}. Shifted population sampling provides coverage over responsiveness distributions, while perturbed capacity trajectories provide coverage over operating regimes.

Together, these two sources of variation train the interface to use the population summary $z_t$ under both composition shifts and changing resource targets. Ablations of the architecture and shifted-sampling components are provided in Appendix~\ref{app:ablations}.

\section{Empirical Evaluation}
\label{sec:experiments}
We evaluate population-aware coordination interfaces along four dimensions: robustness under population composition shift, capacity-control quality, scalable coordination from compact population summaries, and cold-start Sim2Real transfer. For primal maps, we measure forecast fidelity using MAPE. For dual maps, we measure capacity adherence using per-period relative violation $\mathrm{Viol}_t = (J_t - G_t)_+ / G_t$: mean violation on near-limit (NL) periods, the fraction of weeks with violation greater than 10\%, and the fraction of near-limit weeks with violation greater than 10\%. A period is near-limit if the unconstrained policy exceeds 90\% of capacity. Unless otherwise noted, evaluations use 50 inbound-capacity paths over a one-year horizon, with 95\% confidence intervals across paths.

\subsection{Adaptation under Population Composition Shift}
\label{sec:adaptation_experiments}

We first evaluate whether the learned interfaces remain reliable when the active product population changes. We use the demand-decile population shifts defined in Section~\ref{subsec:dist_shift} as a controlled stress test: across the evaluated range of $\alpha$, the target populations span a $30\times$ range in average weekly demand. Because demand is one of the strongest drivers of agent-level cost sensitivity (Appendix~\ref{app:population_shift}), this shift axis creates an extreme perturbation to aggregate response $J_{0:T}$. Natural population drift over time or planner-selected sub-populations would produce milder shifts along this axis; this evaluation is therefore intended as a worst case stress test.

\begin{figure}[htbp]
\centering
\begin{subfigure}{0.47\textwidth}
    \centering
    \includegraphics[width=\textwidth]{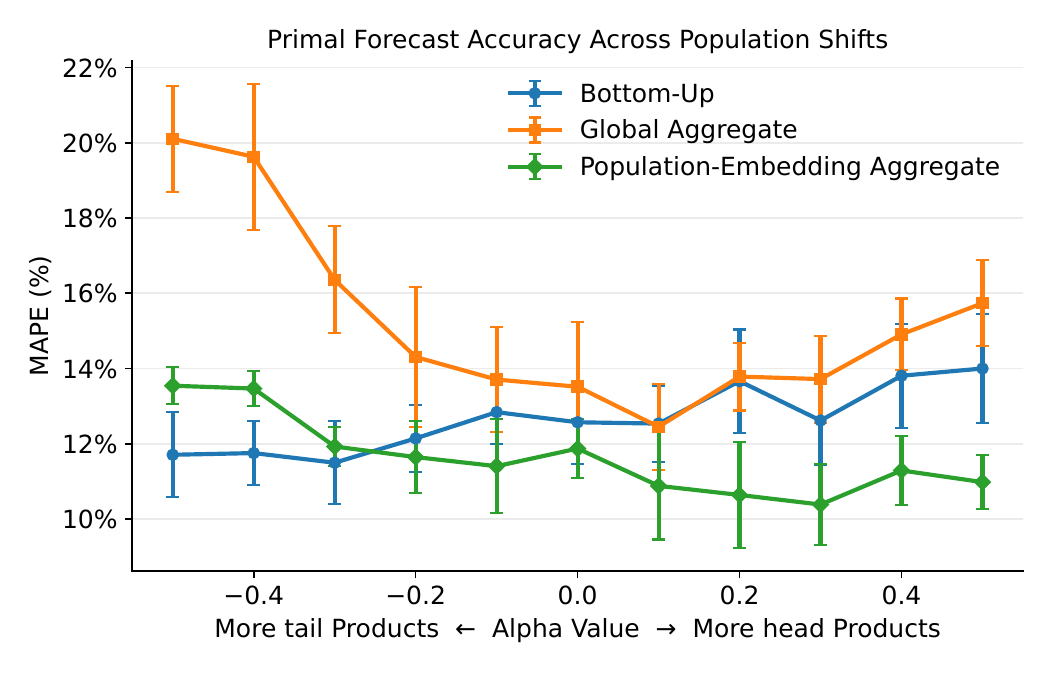}
    \caption{Primal interface fidelity (MAPE).}
    \label{fig:population_shift_primal}
\end{subfigure}
\hfill
\begin{subfigure}{0.47\textwidth}
    \centering
    \includegraphics[width=\textwidth]{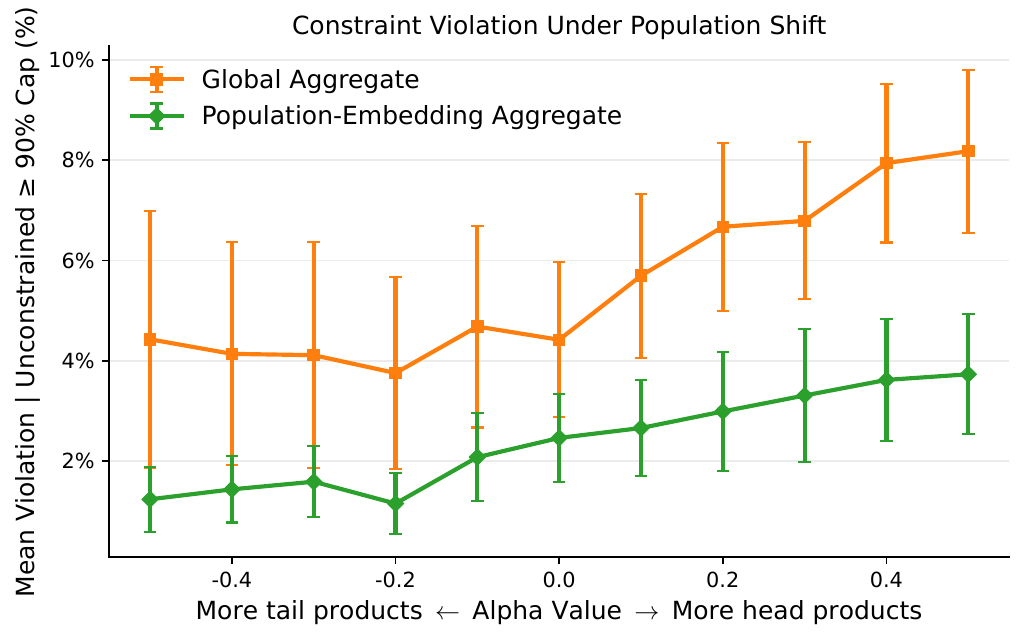}
    \caption{Dual control quality (mean violation).}
    \label{fig:population_shift_dual}
\end{subfigure}
\caption{\small Performance under population shift. Left: aggregate inbound forecast accuracy across shifted populations. Right: capacity-control violation under the corresponding shifted populations. Lower is better in both panels.}
\label{fig:population_shift}
\end{figure}

Figure~\ref{fig:population_shift} reports results for both interface types. The left panel evaluates primal forecast accuracy, and the right panel evaluates dual control quality using mean violation on near-limit periods. Additional shift results and the remaining dual metrics are provided in Appendix~\ref{app:additional_eval}.

Population-aware interfaces are substantially more robust under composition shift than population-unaware baselines. In the primal setting, the Population-Embedding Aggregate maintains roughly 10\% MAPE through moderate shifts ($|\alpha|\leq 0.3$), with noticeable degradation only under more extreme shifts. In the dual setting, the population-aware controller maintains consistently lower near-limit violations across the full shift range.

The baselines degrade more substantially. The Global Aggregate model's MAPE rises from roughly 13\% at $\alpha=0$ to above 20\% near $\alpha=\pm0.5$, and its dual controller incurs consistently higher violations, indicating that aggregate features alone do not preserve the response-relevant composition of the population. The Bottom-Up model also degrades in head-heavy populations: while it captures individual response effects, its product-level training objective and lack of population-level information sharing make it vulnerable to composition-dependent errors, so local errors can accumulate systematically under composition shift.

\subsection{Capacity Control: Primal Search vs. Dual Prediction}
\label{sec:control_experiments}

We isolate the capacity-control subproblem inside the planning loop: given a fixed target plan $G_{0:T}$, can the interface produce cost signals that keep aggregate utilization within capacity? This experiment tests the control primitive the planner calls when evaluating or enforcing a candidate plan. We compare two ways to solve this subproblem: primal iterative search using the forward response map (Appendix~\ref{app:primal_search}) and dual direct prediction, each against the corresponding population-unaware baseline.

Table~\ref{tab:control_main} shows that both design choices matter. Replacing independent per-product forecasts with a learned aggregate primal model improves search: Primal/Embedding reduces mean violation from 8.6\% to 4.2\% and near-limit mean violation from 13.4\% to 7.7\%. Direct dual prediction gives the best adherence, achieving 1.6\% mean violation and 3.1\% near-limit mean violation with the embedding representation. Across both interface directions, population-aware embeddings improve control quality, indicating that effective coordination requires modeling how active-population composition shapes aggregate response.

These results show that the two interface directions serve complementary planner roles: the dual map provides low-latency control for a specified target plan, while the primal map supports search and counterfactual evaluation when the planner needs to reason over candidate cost trajectories.

\begin{table}[htbp]
\centering
\caption{\small Capacity-control adherence across interface types. All entries are percentages (lower is better). Mean and NL-Mean report average percentage capacity violation across all weeks and near-limit weeks, respectively. Viol$_{10}$ and NL-Viol$_{10}$ report the percentage of rollout weeks with violation greater than 10\%, over all weeks and near-limit weeks, respectively. }
\label{tab:control_main}
\small
\setlength{\tabcolsep}{4pt}
\begin{tabular}{lccccc}
\toprule
Method & Mean & NL-Mean & Viol$_{10}$ & NL-Viol$_{10}$ \\
\midrule
Dual / Global Aggregate  & 2.0 & 3.7 & 7.5  & 14.2   \\
Dual / Embedding  & \textbf{1.6} & \textbf{3.1} & \textbf{5.5} & \textbf{10.4}  \\
Primal / Bottom-Up    & 8.6 & 13.4 & 22.0 & 34.1  \\
Primal / Embedding & 4.2 & 7.7  & 11.0 & 18.7  \\
\bottomrule
\end{tabular}
\end{table}

\subsection{Coordination at Scale}
\label{sec:scalability_experiments}

We evaluate whether compact source cohorts can support coordination of larger target populations. For the primal map, the cohort estimates normalized response, which is rescaled to the target-population level via inverse-probability weighting~\citep{sarndal2003model,horvitz1952generalization} (Appendix~\ref{app:primal_scale_handling}). For the dual map, costs inferred from the cohort summary are applied directly to the full target population.

\begin{figure}[htbp]
\centering
\begin{subfigure}{0.47\textwidth}
    \centering
    \includegraphics[width=\textwidth]{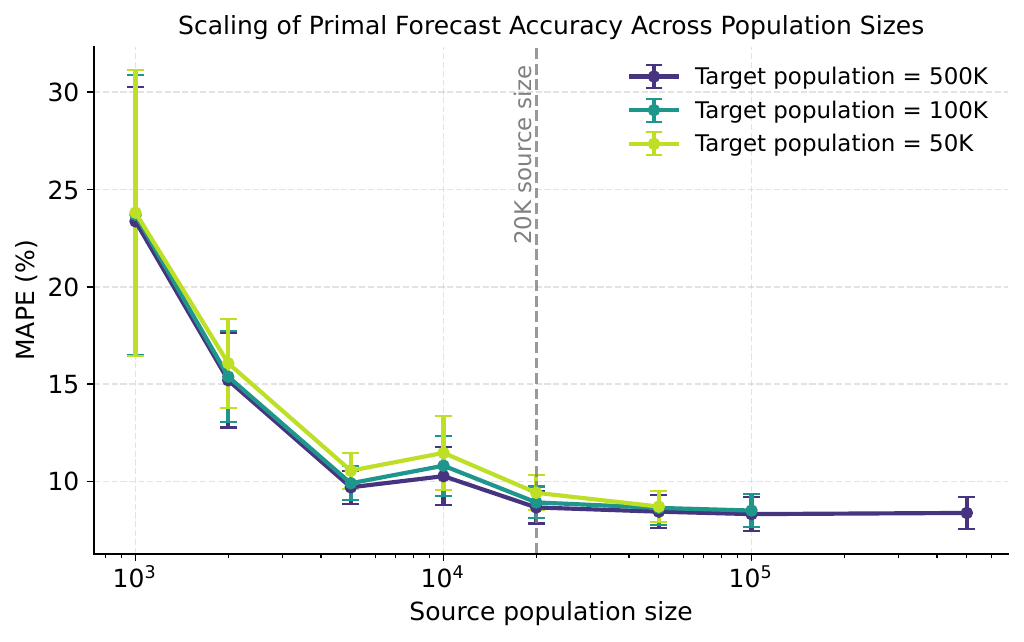}
    \caption{Primal interface fidelity (MAPE).}
    \label{fig:scale_primal}
\end{subfigure}
\hfill
\begin{subfigure}{0.47\textwidth}
    \centering
    \includegraphics[width=\textwidth]{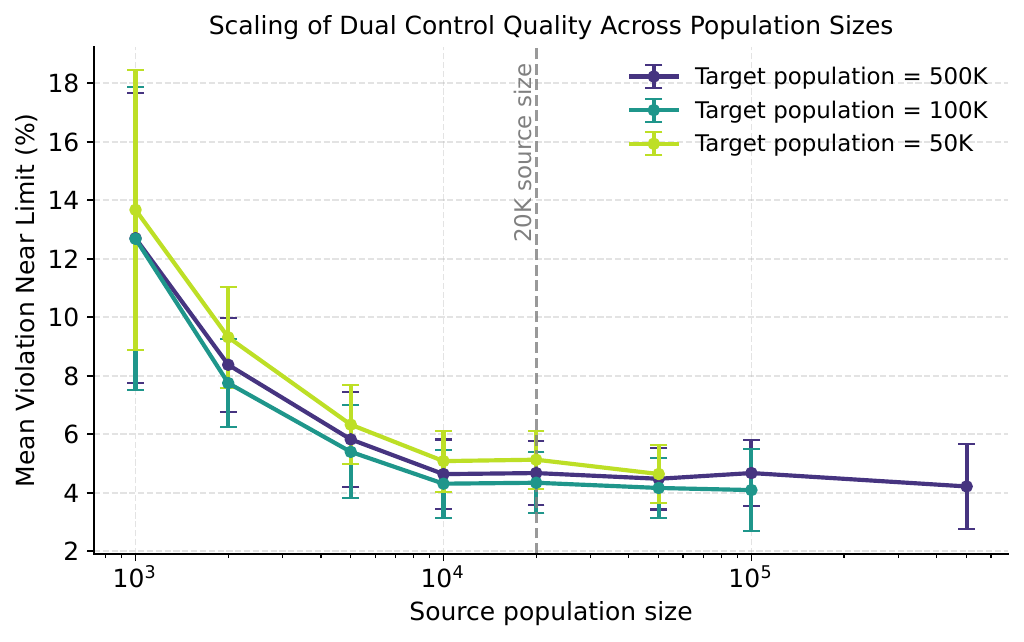}
    \caption{Dual control quality (mean violation).}
    \label{fig:scale_shift_dual}
\end{subfigure}
\caption{\small Coordination interface accuracy under population subsampling from a 500K-agent population. The $x$-axis shows source cohort size on a log scale. Left: primal accuracy (MAPE). Right: dual control quality (mean violation).}
\label{fig:subsampling}
\end{figure}

Figure~\ref{fig:subsampling} shows that performance saturates once the source cohort contains approximately 20K agents. For primal prediction, accuracy at this cohort size is close to full-population inference across target population sizes. For dual control, cost trajectories inferred from 20K-agent cohorts remain effective when applied to substantially larger target populations.

These results show that population-aware interfaces can support prediction and control for large target populations using compact cohorts, without repeated full-population inference.

\subsection{Backtesting Sim2Real Transfer under Cold Start}
\label{subsec:eval_sim2real}

Finally, we evaluate cold-start Sim2Real transfer through a backtest: we train the primal response interface exclusively on simulator-generated rollouts and evaluate it against real aggregate outcomes. We use real observations from a large-scale operational study under two capacity conditions: an unconstrained baseline and constrained operation with a 20\% reduced inbound target. We focus on the primal interface because realized aggregate inbound provides a direct target for evaluation.

As an additional operational reference, we include a product-level Monte Carlo simulation on a comparable-size cohort, but treat it as a directional reference because it is run under a separate policy rather than the exact matched capacity regimes.

Table~\ref{tab:sim2real_nvf} reports the Sim2Real backtest results. The Population-Embedding Aggregate achieves 15.2\% MAPE in the unconstrained setting and 11.1\% MAPE in the constrained setting, outperforming both the Bottom-Up and Global Aggregate baselines. All learned interfaces also perform better relative to the simulation reference, indicating that simulator-trained response interfaces can transfer well to real aggregate outcomes.

We further apply lightweight online calibration using early on-policy observations, fitting a linear regression from simulator-trained forecasts to realized aggregate utilization. This reduces residual bias for the population-aware forecaster and improves its accuracy by 4--6\%. Notably, calibration degrades Bottom-Up performance, likely because its aggregated product-level predictions have a structural mismatch with realized outcomes that a linear correction cannot fix.

Overall, these results show that the learned primal response interface provides aggregate-response predictions that can support planning from day one under cold start and improve further with lightweight online calibration.

\begin{table}[htbp]
\centering
\caption{\small Sim2Real backtest results.  Models are trained on simulated rollouts and evaluated on real observations. Values are MAPE (\%); lower is better. $^\dagger$Simulation Baseline performs product-level Monte Carlo rollouts on a similar real-world cohort under a reference policy; not directly comparable to the evaluated settings.}
\small
\begin{tabular}{lcccc}
\toprule
Method & \multicolumn{2}{c}{Unconstrained} & \multicolumn{2}{c}{Constrained} \\
\cmidrule(lr){2-3}\cmidrule(lr){4-5}
 & Original & +Calib & Original & +Calib \\
\midrule
Bottom-Up                       & 16.4 & 20.1 & 13.3 & 16.2 \\
Global Aggregate                & 22.8 & 21.0 & 18.1 & 13.8 \\
Population-Embedding Aggregate  & \textbf{15.2} & \textbf{9.8}  & \textbf{11.1} &  \textbf{6.9} \\
Simulation Baseline$^\dagger$  & \multicolumn{4}{c}{24.4$^\dagger$} \\
\bottomrule
\end{tabular}
\label{tab:sim2real_nvf}
\end{table}

\section{Conclusion} \label{sec:conclusion}

We introduced population-aware coordination interfaces: learned primal--dual models that expose both a forward aggregate response map and an inverse control map for upstream planners managing large populations of decentralized agents under shared resource constraints. By conditioning on compact, composition-aware population summaries, these interfaces preserve response-relevant heterogeneity while enabling low-latency planner queries.

The empirical results address three challenges that existing approaches do not jointly resolve. First, population-aware interfaces generalize under composition shift, reducing forecast error by 16--19\% and constraint violations by 20--51\% relative to population-unaware baselines. Second, coordination remains scalable as population size grows: compact cohorts of roughly 20K agents support prediction and control for target populations of up to 500K agents in our supply-chain setting. Third, cold-start Sim2Real transfer can be evaluated before real planning through a backtest: simulator-trained forward response models achieve 11.1\% MAPE on real aggregate utilization before on-policy data are available and improve further with lightweight online calibration. Together, these results show that learned response models over compact population summaries can support iterative plan evaluation, direct low-latency capacity control, and backtestable cold-start Sim2Real transfer within a unified coordination abstraction.

\paragraph{Limitations.}
While the coordination-interface abstraction is general, all experiments are in a single supply-chain domain; transfer to other resource-constrained systems remains future work. Scalability results are empirical: we do not characterize theoretically how required cohort size depends on problem complexity. Interface learning relies on a calibrated simulator (evaluated empirically in Appendix~\ref{app:sim2real_ablation}) without theoretical guarantees under mismatch, and our Sim2Real evaluation covers only primal maps; real-world dual evaluation remains future work.

\newpage

\bibliographystyle{ims_nourl_eprint}
\bibliography{external}

\newpage
\appendix
\section{Related Work}
\label{app:related_work}

\paragraph{Multi-Agent Learning and Coordination.}
Centralized-training decentralized-execution methods such as MADDPG~\citep{lowe2017multi} train decentralized policies using centralized information, while Dec-POMDPs formalize decentralized decision-making under partial observability~\citep{oliehoek2016concise}. Value-decomposition methods such as QMIX~\citep{rashid2018qmix} learn monotonic factorizations of centralized action-value functions, and mean-field methods approximate many-agent interactions through aggregate statistics of neighboring agents~\citep{yang2018mean}. Recent work has also studied cooperative MARL for industrial supply-chain settings~\citep{mousa2024analysis}.

Our setting differs from standard MARL policy learning. We do not learn a joint policy, or model direct strategic interaction. Instead, local policies are fixed responders to a broadcast coordination signal. Shared constraints are handled in the dual space: a cost signal makes local decision problems separable, and the learning problem is to model or invert the population-level response to that signal. This makes our approach complementary to MARL: MARL or other methods may train the local policies, while our interfaces coordinate those policies under shared resource constraints.

\paragraph{Dual Decomposition and Learned Capacity Coordination.}
Dual decomposition and shadow pricing are classical tools for coordinating decentralized decisions under shared constraints~\citep{fisher1981lagrangean,bertsekas1999nonlinear,boyd2011distributed}. Related primal--dual and consensus-planning methods decompose decision problems into interacting local and coordinating subproblems~\citep{maggiar2024cpp}. These methods provide principled optimization procedures, but typically require iterative updates of coordination variables rather than amortizing the signal-selection problem into an offline-learned planner-facing interface. Related coordination mechanisms have also been used in supply-chain systems to manage shared capacity and align decentralized inventory decisions~\citep{federgruen1999coordination,cachon2003supply}.

Learning-based inventory systems introduce an additional challenge: evaluating a candidate capacity-cost trajectory may require simulating many product-level policies over a long horizon. Deep RL methods have been applied to large-scale inventory management~\citep{madeka2022deep,gijsbrechts2022can}. The neural coordinator of~\citep{eisenach2024ncc} learns dual cost signals for a given capacity plan, solving the downstream execution problem: given $G_{0:T}$, produce costs that enforce it. Our work addresses the upstream planning problem: providing the planner with reliable counterfactual response maps it can query iteratively to decide \emph{what plan to execute}. This requires both a forward response map (absent in~\citep{eisenach2024ncc}) and population-aware conditioning (since the planning loop runs periodically over evolving populations), in addition to the inverse cost map.

\paragraph{Population Representation and Scalable Aggregation.}
Large-scale coordination requires representing variable and heterogeneous active populations. Hierarchical forecasting methods~\citep{hyndman2011optimal,wickramasuriya2019optimal} address consistency between forecasts at different aggregation levels, but typically assume a fixed hierarchy and focus on reconciliation across levels. Our setting differs because the active population can change across planning cycles, and the model must preserve response-relevant composition rather than only reconcile forecasts across a fixed hierarchy.

Permutation-invariant set encoders such as Deep Sets~\citep{zaheer2017deep} and attention mechanisms~\citep{vaswani2017attention} provide tools for summarizing variable-size collections. We use these ideas to build compact population representations, with a coordination-specific requirement: the summary must preserve heterogeneity in response to a broadcast cost signal, since this heterogeneity determines aggregate utilization under shared constraints. This distinguishes our use of set representations from generic pooling: the representation is evaluated by whether it preserves the response map needed by the planner.

\paragraph{Population Shift and Distribution Robustness.}
Learning under population shift connects to domain adaptation, covariate shift, and dataset-shift literature~\citep{shimodaira2000improving,ben2007analysis,quinonero2009dataset}. Distributionally robust optimization studies learning under uncertainty in the data-generating distribution~\citep{rahimian2019distributionally}, and empirical work shows that subpopulation shifts can degrade performance even when average validation performance is strong~\citep{sagawa2019distributionally,koh2021wilds}. Motivated by this literature, we construct controlled population mixtures along response-relevant attributes to increase training coverage and evaluate robustness under composition shift. Inverse-probability weighting~\citep{shimodaira2000improving} and survey-sampling methods~\citep{sarndal2003model,horvitz1952generalization} provide related tools for correcting nonuniform sampling when estimating population-level quantities; we use related reweighting ideas for cohort construction and scalable evaluation.

\paragraph{Differentiable Simulation and Model Learning.}
Training the inverse cost-prediction model requires differentiating through the simulated response of fixed local policies. Differentiable optimization layers and differentiable simulators have been used to provide gradients for structured prediction, policy optimization, and model-based control~\citep{amos2017optnet,agrawal2019differentiable,suh2022differentiable,parmas2023model}. In inventory settings,~\citep{alvo2023neural} apply differentiable policy optimization through hindsight reformulations. Our use is different: we differentiate through the simulator-induced response of fixed local policies to train a planner-facing inverse cost model, rather than differentiating through a centralized optimization layer or updating the local policies themselves.

\paragraph{Sim2Real Transfer of Forecast Models.}
Sim2Real transfer has been widely studied in robotics and reinforcement learning, where the central challenge is bridging the ``reality gap'' between simulated and physical environments~\citep{jakobi1995evolutionary,tobin2017domain,peng2018sim,tan2018sim}. Domain randomization and dynamics randomization train policies under diverse simulated conditions to improve real-world transfer~\citep{tobin2017domain,peng2018sim}, while other approaches adapt simulation-trained controllers or learned dynamics models using real-world data~\citep{tan2018sim,nagabandi2018neural}. These works primarily transfer agent policies or dynamics models used inside the control loop. Our setting transfers an aggregate-response model trained from simulation: the local policies are fixed, and the transferred model predicts how the population's aggregate utilization responds to capacity-cost signals.

To our knowledge, prior work has not jointly studied the coordination problem through offline-learned planner-facing interfaces that enable forward response prediction and inverse signal prediction, preserve population-composition information for robustness under data shift, scale through compact cohort representations, and support cold-start Sim2Real transfer to real outcomes.

\section{Capacity-Controlled Exogenous IDP} \label{app:exo_idp}
\subsection{Interactive Decision Process Formulation}
\label{app:idp_formulation}

This appendix instantiates the resource-constrained IDP from
Section~\ref{sec:formulation} in the supply-chain simulator used to generate
counterfactual rollout data. We consider discrete time steps $t=0,\ldots,T$,
where $S_t$ is the active product population and each agent $i\in S_t$
corresponds to a product governed by a local replenishment policy. The shared
resource is inbound capacity ($m=1$), so $J_t$, $G_t$, and $\lambda_t$ are
per-period resource-utilization, target-capacity, and
capacity-cost signals. The dual variable $\lambda_t \ge 0$
associated with this constraint is introduced in
Section~\ref{app:constrained_opt} below.

\paragraph{State space.}
The state of each product separates into endogenous inventory variables and
exogenous processes:
\[
x_t^i = (y_t^i,\;\eta_t^i),
\]
where $y_t^i$ denotes endogenous inventory variables and
\[
\eta_t^i = (D_t^i,\;p_t^i,\;c_t^i,\;v_t^i)
\]
is the exogenous state, where $D_t^i$ is the unconstrained demand, $p_t^i$ is sales price,
$c_t^i$ is ordering cost, and $v_t^i$ is vendor lead time. The key Exo-IDP property is that $\eta_{0:T}^i$ evolves
independently of the policy: historical realizations can be replayed under
counterfactual actions while endogenous variables are simulated forward.

\paragraph{Control process.}
At each time $t$, each product observes its local state and selects an order
quantity according to a local replenishment policy:
\[
a_t^i \sim \pi(\cdot \mid x_t^i).
\]

\paragraph{Transition dynamics.}
The endogenous state evolves as
\[
y_{t+1}^i \sim P(\cdot \mid y_t^i,\;a_t^i,\;\eta_t^i).
\]
Orders placed in previous periods arrive during period $t$ according to their
realized lead times:
\[
J_t^i = \sum_{0\le k<t} a_k^i\,\mathbf{1}_{\{v_k^i = t-k\}},
\]
where $v_k^i$ is the realized vendor lead time for the order placed at time $k$.
Beginning-of-period inventory after receipts is
\[
I_{t-}^{i} = I_{t-1}^i + J_t^i.
\]
End-of-period inventory after demand fulfillment is
\[
I_t^i = (I_{t-}^{i} - D_t^i)^+,
\]
and realized sales are
\[
S_t^i = \min(D_t^i,\;I_{t-}^{i}).
\]
Aggregate inbound utilization across the active population is
\[
J_t = \sum_{i\in S_t} J_t^i.
\]

\paragraph{Reward.}
The per-agent utility $U_t^i(x_t^i,a_t^i)$ from Section~\ref{sec:formulation}
is instantiated as a revenue-minus-ordering-cost reward:
\[
R_t^i = p_t^i S_t^i - c_t^i a_t^i.
\]
The unconstrained inventory-control objective is
\[
\max_{\pi}\;\mathbb{E}\!\left[\sum_{t=0}^T \sum_{i\in S_t} R_t^i\right].
\]

\subsection{Constrained Optimization and Lagrangian}
\label{app:constrained_opt}

The planner enforces a per-period inbound capacity limit \(G_t\), requiring
aggregate inbound utilization to satisfy
\[
\mathbb{E}[J_t] \le G_t, \qquad t=0,\ldots,T.
\]
This constraint couples all product-level ordering decisions through the shared
inbound resource. Introducing nonnegative dual variables \(\lambda_t \ge 0\),
the Lagrangian relaxation is
\[
\mathcal{L}(\pi,\lambda)
=
\mathbb{E}\!\left[
\sum_{t=0}^T \sum_{i\in S_t}
\left(R_t^i-\lambda_t J_t^i\right)
\right]
+
\sum_{t=0}^T \lambda_t G_t .
\]
The dual variable \(\lambda_t\) acts as a broadcast cost signal: each product
internalizes the per-unit inbound penalty \(\lambda_t J_t^i\), making the local
policy cost-sensitive:
\[
a_t^i \sim \pi(\cdot \mid x_t^i,\;\lambda_t).
\]
Higher \(\lambda_t\) increases the marginal penalty on inbound utilization,
steering local policies toward reduced ordering; lower \(\lambda_t\) relaxes
this penalty.

At an optimal primal--dual solution, complementary slackness gives
\[
\lambda_t^*\bigl(\mathbb{E}[J_t]-G_t\bigr)=0,
\]
so $\lambda_t^*>0$ only when the constraint binds in expectation.

\subsection{Simulator Construction}
\label{app:gym_construction}

Given historical data $\mathcal{D}$ collected under a historical operating policy
$\pi_{\mathcal{D}}$, we construct a simulator
\[
\mathcal{G} = (\boldsymbol{\eta}_{\mathcal{D}},\;\hat{P},\;\hat{R}),
\]
where $\boldsymbol{\eta}_{\mathcal{D}}$ are replayed historical exogenous
sequences, and $\hat{P}$, $\hat{R}$ are learned approximations of the
endogenous transition dynamics $P$ and per-agent reward $R_t^i$ defined in
Section~\ref{app:idp_formulation}. Conditional on $\boldsymbol{\eta}_{\mathcal{D}}$, endogenous
states are simulated through $\hat{P}$, enabling counterfactual rollouts under
alternative capacity plans and coordination signals.

Following \citep{madeka2022deep}, Exo-IDP simulators constructed from historical
data provide learnability guarantees for counterfactual policy evaluation when exogenous trajectories are replayable and endogenous dynamics are sufficiently calibrated. For the coordination setting here, this requires the simulator to reproduce the policy-conditioned response mapping with enough fidelity to
provide a useful training signal. Appendix~\ref{app:sim2real_ablation} evaluates simulator fidelity under the empirical conditions most relevant to deployment.

\section{Architecture Details} \label{app:architecture}

\subsection{Base Architectures}

The main text writes the interface inputs compactly as
$\psi_\theta(x_t,S_t,\lambda_{t:t+L})$ and
$\phi_\theta(x_t,S_t,G_{t:t+L})$. In this appendix, we expand $x_t$ into
per-product histories, aggregate/global histories, and future known covariates.
For clarity, we describe the architecture using the primal prediction task
$\hat{J}_{t:t+L}$ as the running example; the same encoder and population-summary
construction applies to the dual interface, with $G_{t:t+L}$ replacing
$\lambda_{t:t+L}$ and a cost-output head replacing the utilization-output head.

\paragraph{Input.}
We use $H{=}64$ historical steps and $L{=}26$ forecast horizon.
\begin{itemize}
  \item $x^{i}_{t-H:t} \in \mathbb{R}^{H \times d^i}$: per-product historical features for product $i \in S_t$
  \item $x^{(g)}_{t-H:t} \in \mathbb{R}^{H \times d_g}$: global historical features (aggregated across products)
  \item $x^{(f)}_{t:t+L} \in \mathbb{R}^{L \times d_f}$: future known features (e.g., holidays, forecasted demand)
  \item Capacity signal: $\lambda_{t:t+L} \in \mathbb{R}^{L}_{\ge 0}$ for the primal interface (proposed cost trajectory); $G_{t:t+L} \in \mathbb{R}^{L}_{\ge 0}$ for the dual interface (target capacity plan)
\end{itemize}

\paragraph{Encoder.}
Each historical sequence is mapped to a fixed-size embedding by a stacked dilated
causal CNN:
\[
e_t^i = \mathrm{Enc}_\theta(x^i_{t-H:t}), \quad \text{dilation rates } [1,2,4,8,16,32].
\]
Applied to the global history $x^{(g)}_{t-H:t}$ this yields $e^{(g)}_t$; applied to
each product history $x^i_{t-H:t}$ this yields $e_t^i \in \mathbb{R}^{d_e}$, where
$d_e{=}32$ is the encoder output dimension.

\paragraph{Decoder.}
For primal prediction, the decoder maps the relevant state representation and proposed
cost trajectory to a multi-horizon utilization forecast. The variants differ in how they represent $S_t$: Global Aggregate uses aggregate features,
Bottom-Up sums independent per-agent predictions over $S_t$, and population-aware variants encode
$S_t$ through a population embedding $z_t$; implementation details for $z_t$ are provided in Section~\ref{sec:pooling}.
\[
\hat{J}_{t:t+L} =
\begin{cases}
  \mathrm{Dec}_\theta(e^{(g)}_t,\; x^{(f)}_{t:t+L},\; \lambda_{t:t+L}),
    & \text{Global Aggregate}, \\[4pt]
  \displaystyle\sum_{i\in S_t}\mathrm{Dec}_a(e_t^i,\; x^{(f)}_{t:t+L},\; \lambda_{t:t+L}),
    & \text{Agent Bottom-Up}, \\[4pt]
  \mathrm{Dec}_\theta(z_t,\; x^{(f)}_{t:t+L},\; \lambda_{t:t+L}),
    & \text{Population-Embedding Aggregate}.
\end{cases}
\]
For the dual interface, the same population representation is used with
$G_{t:t+L}$ as input and a cost-output head producing $\hat{\lambda}_{t:t+L}$.
Concretely, each decoder processes the planner-provided trajectory and future features
through another dilated CNN, concatenates the result with the remaining inputs, and applies
a two-layer ELU MLP to produce its horizon-level output.

\paragraph{Scale handling for the primal population embedding.}
\label{app:primal_scale_handling}

The main text writes the Population-Embedding primal decoder compactly as
\[
\hat{J}_{t:t+L}
=
\mathrm{Dec}^{\mathrm{primal}}_\theta(z_t,\lambda_{t:t+L}).
\]
This denotes the final aggregate-utilization output of the primal head. In the actual implementation, because \(z_t\) captures response-relevant composition rather than absolute population size, the neural decoder first predicts normalized utilization,
\[
\widehat{r}_{t:t+L}
=
\widetilde{\mathrm{Dec}}^{\mathrm{primal}}_\theta
\left(z_t,\lambda_{t:t+L}\right),
\]
and converts it to absolute aggregate utilization using a population-level scale:
\[
\widehat{J}_{t:t+L}
=
\widehat{r}_{t:t+L}D(S_t),
\qquad
D(S_t)=\sum_{i\in S_t}d^i.
\]
In our supply-chain experiments, \(D(S_t)\) is the aggregate historical demand level of the population, and \(d^i\) is a scalar historical demand-level estimate for product \(i\).

In the scalability experiments, a source cohort \(S \subseteq \mathcal{A}\) is sampled from a larger target population \(\mathcal{A}\). The cohort predicts the normalized response \(\widehat{r}_{t:t+L}(S)\), while the target-population demand scale is estimated by the Horvitz--Thompson estimator~\citep{horvitz1952generalization,sarndal2003model}:
\[
\widehat{D}(\mathcal{A})
=
\sum_{i\in S}\frac{d^i}{p_i},
\qquad
p_i=\Pr(i\in S).
\]
This estimator is unbiased for the target-population demand total \(D(\mathcal{A})=\sum_{i\in\mathcal{A}}d^i\). For uniform sampling with target population size \(N\) and cohort size \(n\), \(p_i=n/N\), so
\[
\widehat{D}(\mathcal{A})
=
\frac{N}{n}\sum_{i\in S}d^i.
\]
The target-population primal forecast is then
\[
\widehat{J}_{t:t+L}(\mathcal{A})
=
\widehat{r}_{t:t+L}(S)\widehat{D}(\mathcal{A}).
\]

Thus, inverse-probability weighting is applied to the demand scale used to convert normalized utilization to absolute utilization. The remaining approximation is whether the sampled cohort representation yields a normalized response close to the full-population normalized response; this is evaluated empirically in Section~\ref{sec:scalability_experiments}. For the dual interface, no demand scaling is applied because the output is a shared broadcast cost trajectory applied directly to the target population.

\subsection{Population Pooling Variants} \label{sec:pooling}

We now specify how the population-aware variants construct the embedding $z_t$.

\paragraph{Population-Embedding (per-Agent).}
Per-product embeddings $e_t^i=\mathrm{Enc}_\theta(x^i_{t-H:t})$ are aggregated by
attention. For the primal interface, the attention queries are constructed from the
proposed cost trajectory and future known features:
\[
q_{1:L}
=
\mathrm{CNN}_{\mathrm{query}}
\left(
\lambda_{t:t+L}
\oplus x^{(f)}_{t:t+L}
\right),
\]
where $x^{(f)}_{t:t+L}$ denotes global future features. For the dual interface,
$G_{t:t+L}$ replaces $\lambda_{t:t+L}$ in the query input.

These embeddings and queries are projected to queries, keys, and values via
learned weight matrices with layer normalization ($\mathrm{LN}$);
for each horizon step $\ell=1,\ldots,L$:
\[
Q_\ell=W_Q\mathrm{LN}(q_\ell),\qquad
K^i=W_K\mathrm{LN}(e_t^i),\qquad
V^i=W_V\mathrm{LN}(e_t^i).
\]
Attention over products gives
\begin{align}
  \alpha^i_\ell
  &=
  \mathrm{softmax}_i\!\left(
  \frac{Q_\ell {K^i}^\top}{\sqrt{d_k}}
  \right),
  \qquad
  \tilde{V}_\ell
  =
  \sum_{i\in S_t}\alpha^i_\ell V^i, \notag\\
  \qquad
  \beta_\ell
  &=
  \mathrm{softmax}_\ell(w^\top \tilde{V}_\ell),\qquad
  z_t
  =
  \sum_{\ell=1}^L \beta_\ell \tilde{V}_\ell.
  \label{eq:att}
\end{align}

Thus, in primal mode,
$z_t=\rho_{\mathrm{att}}(\{e_t^i\}_{i\in S_t},\lambda_{t:t+L},x^{(f)}_{t:t+L})$.

\paragraph{Population-Embedding (Bucketized) Aggregate}
The bucketized variant groups products into $K{=}10$ demand buckets before attention,
reducing per-product noise while preserving composition information.

\textbf{Stage 1: within-bucket attention pooling.}
Let \(\mathcal{B}_k \subseteq S_t\) be the set of products in bucket \(k\), and let
\[
d^i=\frac{1}{H}\sum_{\tau=t-H}^{t-1}\hat{D}^i_\tau
\]
denote the mean historical demand of product \(i\) over \([t-H,t)\). For each bucket, the bucket embedding is computed by applying the same attention mechanism as Eq.~\eqref{eq:att}, restricted to products in that bucket:
\[
e^{(b)}_{k,t}
=
\rho_{\mathrm{att}}\!\left(
\{e_t^i\}_{i\in\mathcal{B}_k},
\lambda_{t:t+L},
x^{(f)}_{t:t+L}
\right).
\]
We also append bucket prevalence features that encode the bucket's count share and demand share:
\begin{align}
p^{(\mathrm{count})}_k
&=
\frac{|\mathcal{B}_k|}
{\sum_{\ell=1}^K|\mathcal{B}_\ell|},
\qquad
p^{(\mathrm{demand})}_k
=
\frac{\sum_{i\in\mathcal{B}_k} d^i}
{\sum_{i\in S_t} d^i}, \notag\\
\tilde{e}^{(b)}_{k,t}
&=
e^{(b)}_{k,t}
\oplus
\bigl[p^{(\mathrm{count})}_k,\ p^{(\mathrm{demand})}_k\bigr].
\label{eq:bucket}
\end{align}

\textbf{Stage 2: bucket-level attention and decoding.}
We apply the same attention mechanism as Eq.~\eqref{eq:att} over the $K$ bucket
embeddings $\{\tilde{e}^{(b)}_{k,t}\}_{k=1}^K$ to obtain the population embedding:
\[
z_t =
\rho_{\mathrm{bucket}}\!\bigl(
\{\tilde{e}^{(b)}_{k,t}\}_{k=1}^K,\lambda_{t:t+L},x^{(f)}_{t:t+L}
\bigr).
\]
The resulting $z_t$ is passed to the same decoder form as the per-agent population
embedding.

\subsection{Coordinator Input Features}

The coordination interfaces consume the following features.
Historical features span $[t{-}H, t)$; future features span $[t, t{+}L)$.

\begin{itemize}
  \item Historical state (aggregate, $[t{-}H,\,t)$):
    order quantities $\sum_i a^i$,
    inventory $\sum_i I^i$,
    availability-corrected demand $\sum_i \hat{D}^i$,
    inbound $\sum_i J^i$.
  \item Historical state (per-product $i \in S_t$, $[t{-}H,\,t)$):
    order quantity $a^i$, inventory $I^i$, demand $\hat{D}^i$, inbound $J^i$,
    unit economics (price, cost).
  \item Forecasted aggregate ($[t,\,t{+}L)$, $L$ weeks):
    mean demand $\hat{D}_{t:t+L} = \sum_i \hat{D}^i_{t:t+L}$;
    projected inventory after expected drain
    $\sum_i\!\bigl(I^i_t - \textstyle\sum_{\tau=t}^{t+L-1} \hat{D}^i_\tau\bigr)$.
  \item Time-series (global): distance to public holidays.
  \item Capacity cost (global):
    proposed cost trajectory $\lambda_{t:t+L}$ (primal interface)
    or target capacity plan $G_{t:t+L}$ (dual interface).
\end{itemize}

\section{Training Implementation} \label{app:training}
\subsection{Capacity-Plan and Cost-Trajectory Generation} \label{app:capacity_curve_sampling}

Training and evaluation require exogenous trajectories that define the coordination problem.
Depending on the interface, the exogenous input is either a target capacity plan
$G_{0:T}$ or a cost trajectory $\lambda_{0:T}$. The dual interface is conditioned on
sampled capacity plans and learns to produce costs; the primal interface is trained on
cost trajectories and the aggregate utilization they induce.

\begin{figure}[htbp]
\centering
\includegraphics[width=0.8\textwidth]{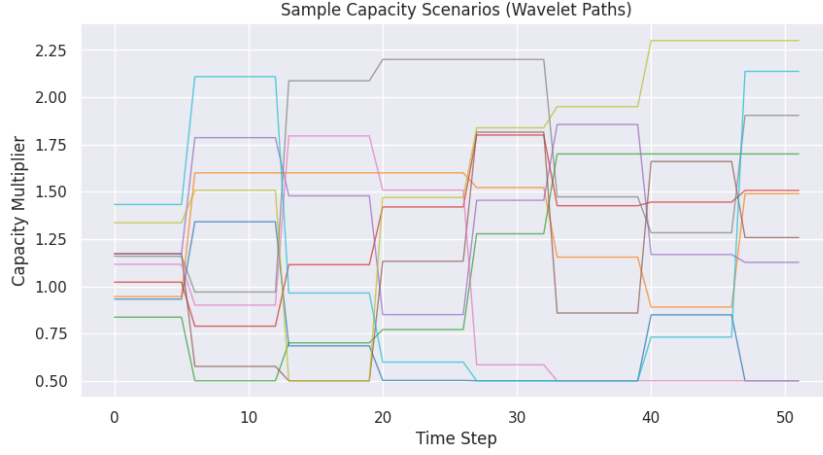}
\caption{\small Example capacity target trajectories generated by the wavelet sampler using a truncated Haar wavelet basis.}
\label{fig:wavelet_sampler}
\end{figure}

\paragraph{Capacity target generation.}
We generate target capacity trajectories $G^{(n)}_{0:T}$ using a truncated Haar wavelet
basis, following~\citep{eisenach2024ncc}. Random Gaussian coefficients produce a diverse
family of piecewise-constant capacity paths with controlled temporal variation, capturing
realistic on/off capacity shifts (Fig.~\ref{fig:wavelet_sampler}). Each sampled path is
rescaled to the aggregate utilization scale of the target population; in our supply chain
setting, this corresponds to aggregate inbound volume.

\paragraph{Coordination signal generation.}
Cost trajectories for primal training can be generated from several sources, including
random priors over $\lambda$-space, historical production costs, or learned capacity
controllers. We use a constraint-space sampling strategy: first sample target capacity
plans $G_{0:T}\sim\mathcal{P}_G$, then map them to broadcast cost trajectories using a
trained dual coordinator. This focuses the primal training distribution on cost signals
that are relevant to feasible capacity-control problems, rather than arbitrary perturbations
in $\lambda$-space.

Given a sampled target $G^{(n)}_{0:T}$, the trained dual coordinator is
applied step by step to produce the episode-level cost trajectory
$\lambda^{(n)}_{0:T}$; at each step $t$,
\[
\lambda^{(n)}_{t:t+L}
=
\phi_\theta(x_t^{(n)},\,S_t^{(n)},\,G^{(n)}_{t:t+L}).
\]

The simulator is then rolled out under $\lambda^{(n)}_{0:T}$, applying the broadcast
costs to the fixed local policies and recording the induced aggregate utilization
$J^{(n)}_{0:T}$. This yields supervised primal training pairs
\[
(x_t^{(n)},\,S_t^{(n)},\,\lambda^{(n)}_{0:T}) \;\mapsto\; J^{(n)}_{0:T}.
\]
At training step $t$, the window $\lambda^{(n)}_{t:t+L}$ and the corresponding
target $J^{(n)}_{t:t+L}$ are extracted from each rollout.

For dual training, at each step $t$ the window $G^{(n)}_{t:t+L}$ is extracted
from the sampled trajectory $G^{(n)}_{0:T}$ and used directly in the
closed-loop Direct-Backprop procedure described below.

\subsection{Training Procedures} \label{app:implementation}

The two interfaces have different training procedures. At each training epoch, a capacity-target path $G_{0:T}\sim\mathcal{P}_G$ is
sampled from the wavelet distribution, and a shifted population $S_t$ is drawn via the $\alpha$-parameterized composition shift
(Appendix~\ref{app:population_shift}).

\paragraph{Primal interface: offline supervised MSE.}
The primal response forecaster $\psi_\theta$ is trained independently, without backpropagating
through the local policies. Training data consists of simulator rollouts: for each cost
trajectory $\lambda_{0:T}$, the fixed local policies respond and produce aggregate inbound
$J_{0:T}$, which serves as the supervised target. The model is trained with
\[
\hat{J}_{t:t+L}
=
\psi_\theta(x_t,S_t,\lambda_{t:t+L}),
\qquad
\mathcal{L}_{\mathrm{primal}}(\theta)
=
\sum_t
\left\|
\hat{J}_{t:t+L}-J_{t:t+L}
\right\|_2^2 .
\]
No gradient is propagated through the local policies during primal training.

\paragraph{Dual interface: closed-loop Direct-Backprop.}
In the dual interface, the dual cost predictor $\phi_\theta$ maps a target capacity plan to a
broadcast cost trajectory,
\[
\hat{\lambda}_{t:t+L}
=
\phi_\theta(x_t,S_t,G_{t:t+L}).
\]
Training proceeds via Direct-Backprop~\citep{madeka2022deep,eisenach2024ncc} through the
closed feedback loop:
\begin{enumerate}
  \item A capacity path $G_{0:T}\sim\mathcal{P}_G$ is sampled from the truncated Haar
        wavelet distribution.
  \item $\phi_\theta$ predicts a cost trajectory
        $\hat{\lambda}_{t:t+L}=\phi_\theta(x_t,S_t,G_{t:t+L})$.
  \item The fixed local policies respond to $\hat{\lambda}_{t:t+L}$ in the differentiable
        Exo-IDP simulator, producing simulated aggregate inbound ${J}_t$.
  \item Gradients flow through the simulator response to update $\phi_\theta$ by minimizing
        Eq.~\eqref{eq:coordinator_loss}.
\end{enumerate}

\begin{equation}
  \mathcal{L}_{\mathrm{dual}}(\theta)
  =
  \alpha_{\mathrm{quad}}\sum_{t > t_{\mathrm{burn}}}\bigl({J}_t-G_t\bigr)_+^2
  +
  \alpha_{\ell_1}\sum_t\|\hat{\lambda}_t\|_1
  +
  \alpha_{\mathrm{mse}}\,\mathcal{L}_{\mathrm{mse}},
  \label{eq:coordinator_loss}
\end{equation}
where $(u)_+=\max(u,0)$, and the capacity-violation sum is restricted to
steps after a burn-in of 6 to exclude simulator warm-up.

$\mathcal{L}_{\mathrm{mse}}$ is a forecast-consistency regularizer that penalizes
disagreement between a differentiable surrogate of the aggregate inbound response
and the realized aggregate inbound from the closed-loop rollout:
\[
  \mathcal{L}_{\mathrm{mse}}
  =
  \sum_t \bigl\|\hat{J}_t^{\mathrm{soft}} - J_t\bigr\|_2^2.
\]
Here $J_t$ is the realized aggregate inbound produced by the buying
policies in the closed-loop simulator under the proposed cost trajectory
$\hat{\lambda}_{t:t+L}$; it is recorded as a fixed target and detached from the
computational graph. The surrogate $\hat{J}_t^{\mathrm{soft}}$ is obtained by
running the same buying policies in gradient-enabled mode under
$\hat{\lambda}_{t:t+L}$, allowing gradients to flow back to $\phi_\theta$.
This surrogate is computed entirely within the simulator---there is no separate
output head on $\phi_\theta$. Thus, the regularizer aligns the differentiable
surrogate used for training with the realized simulator trajectory.
The local policies are frozen throughout; only $\phi_\theta$ is updated.

\subsection{Training Hyperparameters} \label{app:hyperparams}

Table~\ref{tab:hyperparams} lists all training hyperparameters for both the
primal and dual interface coordinators. Both interfaces share the same
optimizer, learning rate, batch size, and architecture configuration.

\begin{table}[htbp]
\centering
\small
\caption{Training hyperparameters for the dual interface ($\phi_\theta$, cost predictor)
and the primal interface ($\psi_\theta$, response forecaster).
Values are shared between the two unless otherwise noted.}
\label{tab:hyperparams}
\begin{tabular}{lll}
\toprule
\textbf{Hyperparameter} & \textbf{Value} & \textbf{Notes} \\
\midrule
\multicolumn{3}{l}{\textit{Optimization}} \\
Optimizer & Adam & \\
Learning rate & $10^{-3}$ & Fixed throughout \\
Batch size & 3{,}000 & Population size $|S_t|$ per rollout\\
Training epochs & 2{,}000 & \\
Gradient clipping & 0.01 ($\ell_\infty$-norm) &  \\
\midrule
\multicolumn{3}{l}{\textit{Sequence lengths}} \\
History window $H$ & 64 & Weeks of lookback \\
Forecast horizon $L$ & 26 & Weeks ahead \\
Training episode length   & 156  & Weeks per episode; start randomized \\
Evaluation horizon  &  52  &  Weeks per evaluation rollout; held out from training \\
Burn-in steps & 6 & Steps excluded from loss \\
\midrule
\multicolumn{3}{l}{\textit{Architecture}} \\
CNN channel width & 32 & Both history and future encoders \\
Future CNN channel width & 16 & \\
Decoder hidden size & 32 & Two-layer ELU MLP \\
Attention dimension $d_k$ & 20 & Query/key projection size \\
History dilation rates & $[1,2,4,8,16,32]$ & Stacked causal CNN \\
Future dilation rates & $[1,2,4,8,11]$ & Reversed causal CNN \\
Encoder embedding $d_e$  &  32  &  Per-product encoder output \\
Demand buckets $K$ & 10 & Number of buckets for Bucketized variant  \\
\midrule
\multicolumn{3}{l}{\textit{Training (composition shift)}} \\
Shift quantile buckets  &  10  &  Demand-decile partition for $\alpha$-shift construction \\
Shift values $\alpha$   &  $[-0.5, 0.5]$  &  Population mixtures per epoch \\
\midrule
\multicolumn{3}{l}{\textit{Dual loss coefficients (Eq.~\ref{eq:coordinator_loss})}} \\
$\alpha_{\mathrm{quad}}$ & 100 & Capacity-violation penalty \\
$\alpha_{\ell_1}$ & 1 & Cost sparsity penalty \\
$\alpha_{\mathrm{mse}}$ & 100 & Forecast-consistency penalty \\
\bottomrule
\end{tabular}
\end{table}

\subsection{Primal Cost-Search Procedure}
\label{app:primal_search}

For experiments that use the primal interface inside an iterative cost-search
loop, we optimize a horizon-level cost vector directly while keeping the primal
interface fixed. Let $\tilde{\lambda}_{t:t+L}$ denote unconstrained cost
parameters. We parameterize nonnegative costs as
\[
\lambda_{t:t+L}=\mathrm{ReLU}(\tilde{\lambda}_{t:t+L}),
\]
with $\tilde{\lambda}_{t:t+L}$ initialized to zero.

At each iteration, the primal interface predicts aggregate utilization
\[
\hat{J}_{t:t+L}=\psi_\theta(x_t,S_t,\lambda_{t:t+L}),
\]
and gradients are backpropagated through the frozen primal model to update
the cost parameters $\tilde{\lambda}_{t:t+L}$. The search minimizes the
capacity-violation loss
\[
\mathcal{L}_{\mathrm{search}}
=
\frac{1}{L}\sum_{\ell=0}^{L-1}
\left(\max\left\{0,\hat{J}_{t+\ell}-G_{t+\ell}\right\}\right)^2 .
\]
We optimize $\tilde{\lambda}_{t:t+L}$ using Adam with learning rate $0.1$ for up
to $100$ iterations at each planning step. Search stops early if the mean
absolute percentage error between predicted utilization and the capacity target
falls below $3\%$.

\section{Controlled Population Shift} \label{app:population_shift}
\subsection*{Construction of Controlled Population Shifts}

Section~\ref{subsec:population_response} defines population shift as a change in the
mixture of agent types in the active population $S_t$ that changes the aggregate
response to a fixed broadcast signal. In particular, if agents differ in their
responsiveness to the coordination signal $\lambda_t$, then changing the distribution
of those agents changes the response map $\mathcal{J}_t(x_t,S_t,\lambda_t)$. This appendix
describes the controlled shift construction used both to generate training coverage
and to evaluate robustness under composition shift.

We construct shifted populations by reweighting agents along a response-relevant
attribute $\xi^i$, computed from a historical reference window of the agent's
state. Agents are sorted by $\xi^i$ and partitioned into $K$ quantile buckets
$\{\mathcal{B}_k\}_{k=1}^K$. Let $u_k$ denote the baseline mass of bucket
$\mathcal{B}_k$, and let $\bar{\xi}_k > 0$ denote the bucket-level mean of the
shift attribute. For a shift parameter $\alpha$, we define the bucket sampling
weights
\[
p_k(\alpha)
=
\frac{u_k(\bar{\xi}_k)^\alpha}
{\sum_{\ell=1}^{K}u_\ell(\bar{\xi}_\ell)^\alpha},
\qquad k=1,\ldots,K.
\]
Here $\alpha=0$ recovers the baseline population distribution, $\alpha>0$
upweights buckets with larger $\bar{\xi}_k$, and $\alpha<0$ upweights buckets
with smaller $\bar{\xi}_k$.

Within each bucket $\mathcal{B}_k$, agents are sampled according to a
within-bucket distribution $r_k(i)$, with
$\sum_{i\in\mathcal{B}_k} r_k(i)=1$. The resulting agent-level distribution used to sample shifted populations is
\[
p_\alpha(i)
=
p_{k(i)}(\alpha)\, r_{k(i)}(i).
\]
where $k(i)$ denotes the bucket index of agent $i$. Varying $\alpha$ therefore
produces a controlled family of populations whose composition changes
systematically along the selected attribute.

This construction is used in two places. During training, shifted populations
provide coverage over different responsiveness distributions, allowing the
interface to learn how aggregate response changes with $S_t$. During evaluation,
the same $\alpha$-indexed family defines target populations for stress-testing
generalization under controlled composition shift.

\subsection*{Choice of Shift Dimensions}

We identify response-relevant attributes for the controlled shift by estimating
agent-level cost sensitivity and relating it to observable agent features.

\begin{figure}[htbp]
\centering
\includegraphics[width=0.7\textwidth]{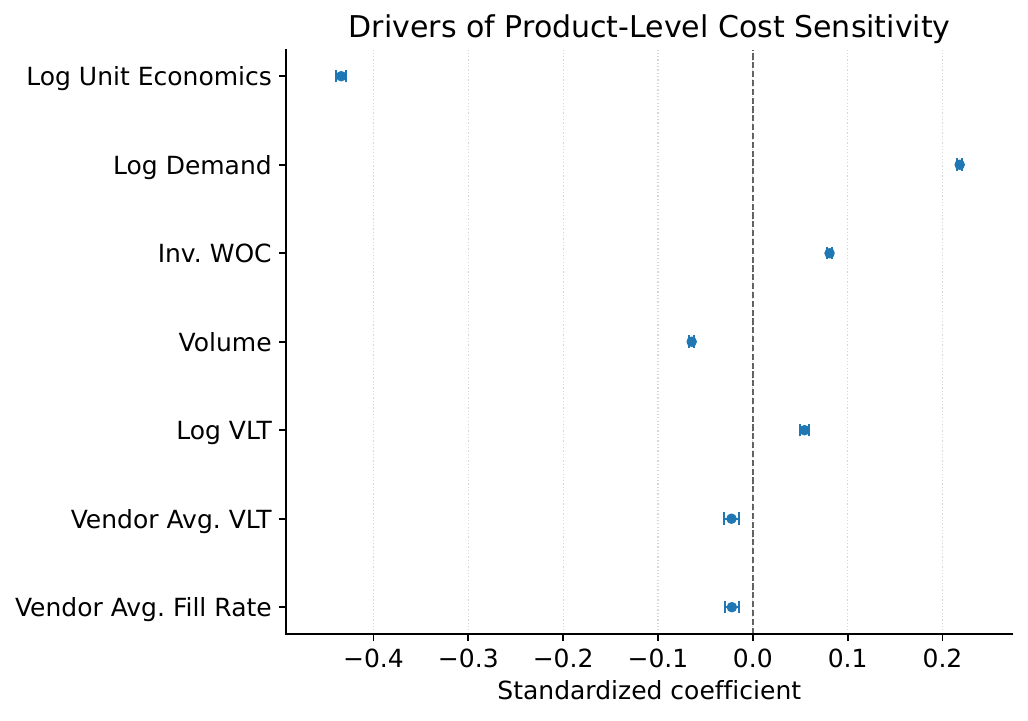}
\caption{Standardized OLS coefficients relating observable product attributes to
estimated product-level cost sensitivity. Whiskers show 95\% confidence
intervals. Larger absolute coefficients indicate stronger association with
responsiveness to the broadcast cost signal.}
\label{fig:sensitivity_reg}
\end{figure}

We use the agent-level bottom-up response model from
Section~\ref{subsec:architecture} to estimate per-agent utilization under a cost
trajectory. Because the model is differentiable with respect to the cost trajectory,
we define a scalar responsiveness score for each agent by averaging local absolute
sensitivities over rollout states and forecast horizons:
\[
s_i
=
\frac{1}{|\mathcal{T}|L}
\sum_{t\in\mathcal{T}}
\sum_{\ell=0}^{L-1}
\left|
\frac{\partial \hat{J}_{t+\ell \mid t}^i}
{\partial \lambda_{t+\ell}}
\right|.
\]
We then regress standardized responsiveness scores on standardized candidate
attributes:
\[
\tilde{s}_i
=
\beta_0+\beta^\top\tilde{q}_i+\varepsilon_i.
\]
In the supply chain instantiation, $\tilde{q}_i$ includes product-level attributes
such as unit economics, demand, inventory coverage (weeks of cover), product volume,
vendor lead time, and vendor fill rate. Demand and unit economics have
the strongest associations with estimated cost sensitivity
(Fig.~\ref{fig:sensitivity_reg}), motivating the controlled shifts used in our
experiments.

\subsection*{Composition of Shifted Populations}
We apply the shift construction with $K{=}10$ quantile buckets along two
dimensions — demand and unit economics — each producing an independent $\alpha$-indexed family of
populations.

\begin{figure}[htbp]
\centering
\includegraphics[width=0.6\textwidth]{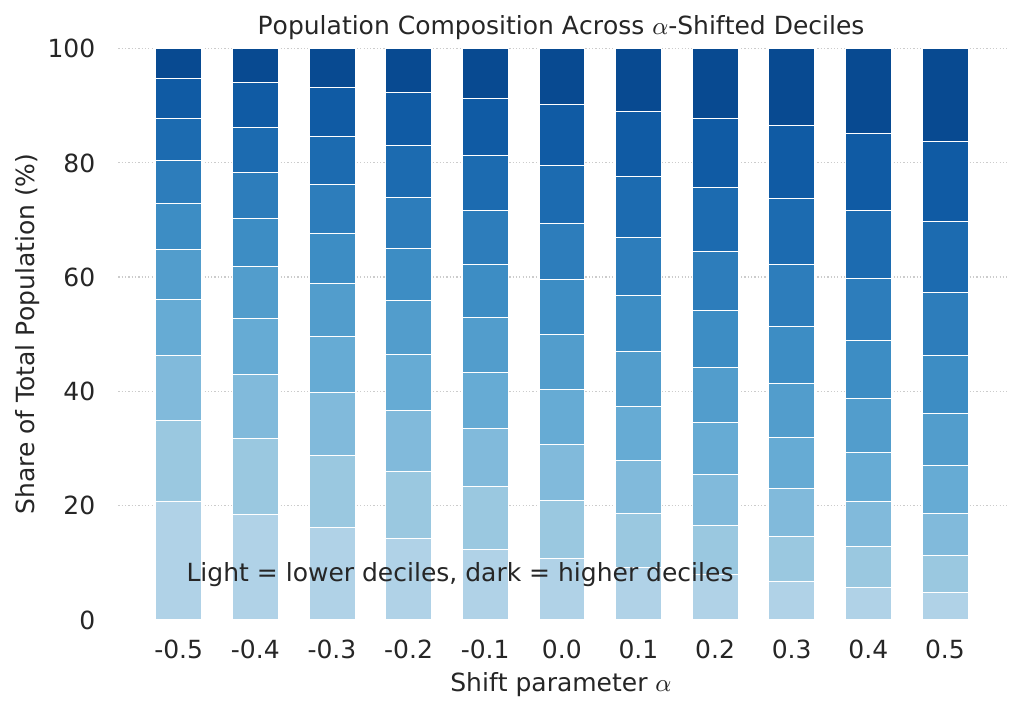}
\caption{\small Composition of target populations under $\alpha$-shifted
distributions, measured as the expected sampling mass assigned to each decile of
the bucketization attribute. Positive $\alpha$ shifts mass toward higher-value
segments; negative $\alpha$ shifts mass toward lower-value segments.}
\label{fig:composition_shift}
\end{figure}

For each value of $\alpha$, we first compute the expected sampling mass assigned
to each attribute decile under $p_\alpha$. Figure~\ref{fig:composition_shift}
visualizes this decile-level mass allocation. Positive $\alpha$ shifts sampling
mass toward higher-value buckets, while negative $\alpha$ shifts mass toward
lower-value buckets.

\begin{figure}[htbp]
\centering
\begin{subfigure}{0.48\textwidth}
    \centering
    \includegraphics[width=\textwidth]{new_figures/dmd_shift_deciles.pdf}
    \caption{Demand deciles}
    \label{fig:asin_percentage_demand}
\end{subfigure}
\hfill
\begin{subfigure}{0.48\textwidth}
    \centering
    \includegraphics[width=\textwidth]{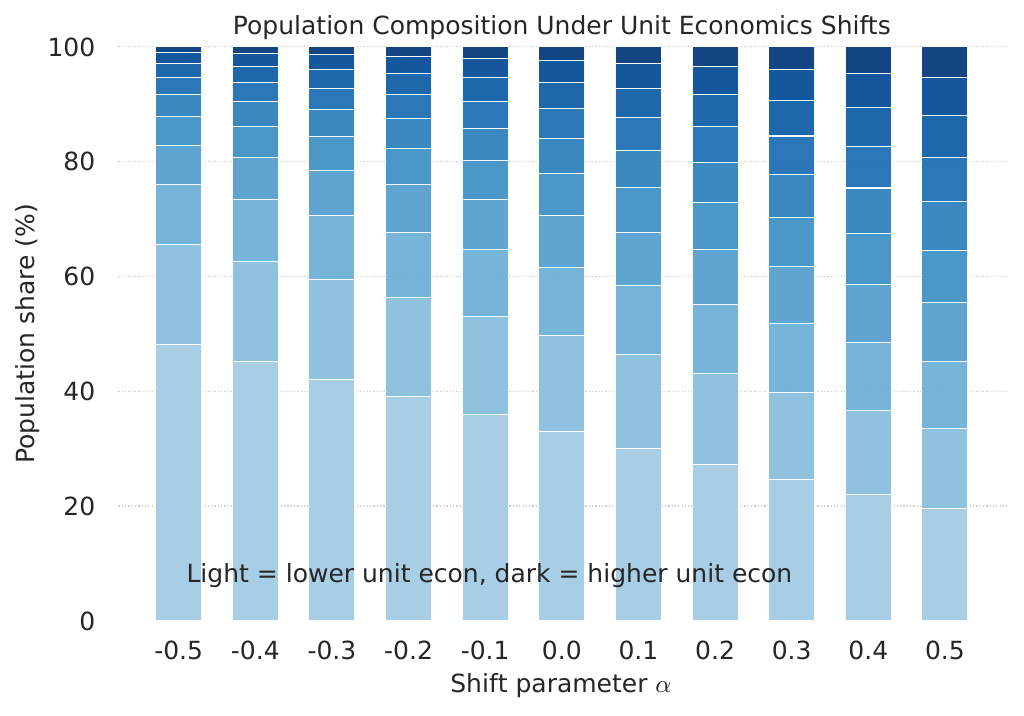}
    \caption{Unit economics deciles}
    \label{fig:asin_percentage_unit_econ}
\end{subfigure}
\caption{\small Realized product-count share in each decile under
$\alpha$-shifted population sampling, for demand (left) and unit
economics (right). The plots show how reweighting by demand or unit economics
changes the product composition of the sampled population.}
\label{fig:asin_percentage}
\end{figure}

Because supply-chain attributes such as demand and unit economics are
right-skewed, expected sampling mass and realized product-count share can differ:
high-value deciles may represent a small fraction of products but a large share of
demand or economic mass. Figure~\ref{fig:asin_percentage} reports the realized
product-count share in each decile for shifted populations constructed along
demand and unit economics.

\subsection*{Effect on Evaluation Populations}
\begin{figure}[htbp]
\centering
\begin{subfigure}{0.48\textwidth}
    \centering
    \includegraphics[width=\textwidth]{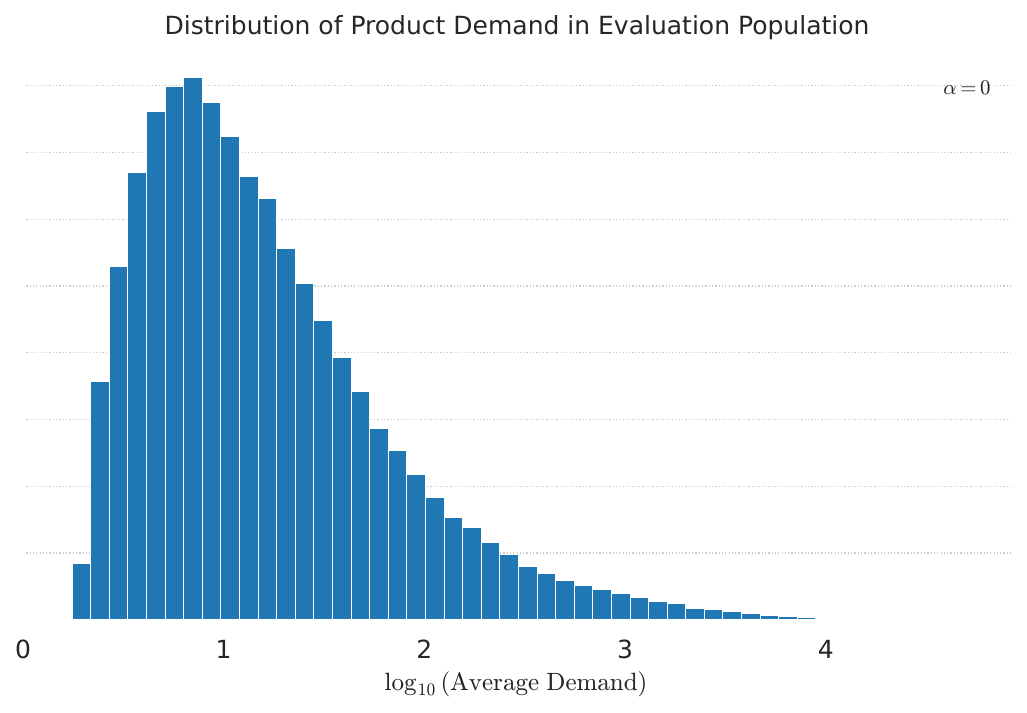}
\end{subfigure}
\hfill
\begin{subfigure}{0.48\textwidth}
    \centering
    \includegraphics[width=\textwidth]{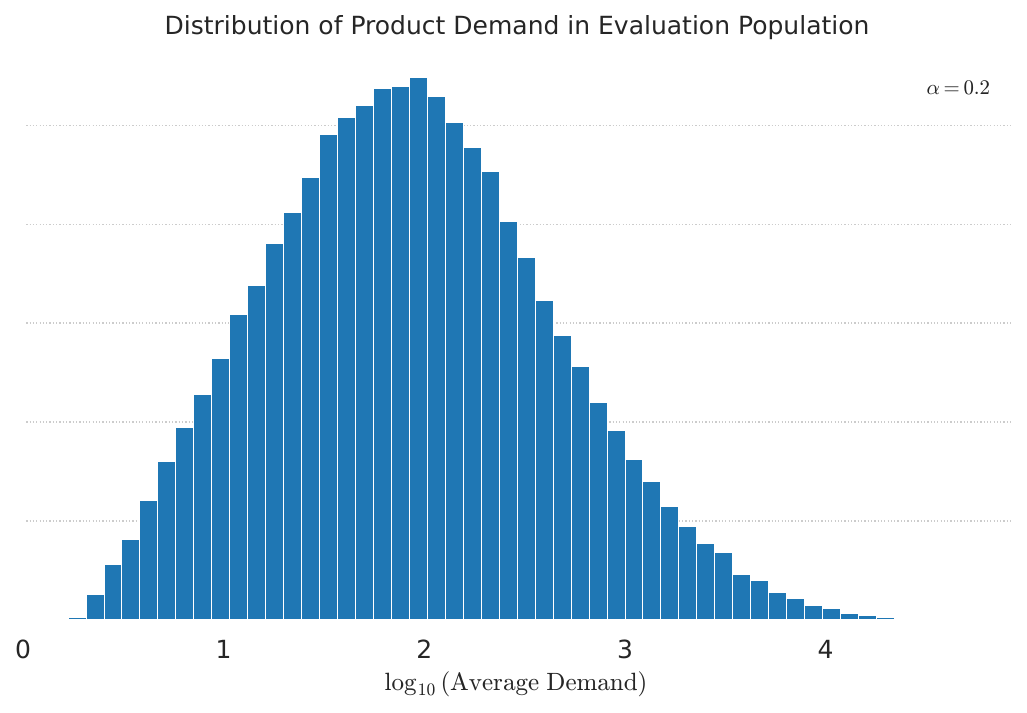}
\end{subfigure}
\caption{\small Product distribution within evaluation populations for $\alpha = 0$
(left, baseline) and $\alpha = 0.2$ (right, shifted), illustrating the reweighting
of population composition toward higher-value segments as $\alpha$ increases.}
\label{fig:demand_histograms}
\end{figure}

In our population-shift evaluation (Section~\ref{sec:adaptation_experiments}), we consider values of $\alpha$ ranging from $-0.5$ to $0.5$. To illustrate the effect of these composition shifts, Figure~\ref{fig:demand_histograms} shows histograms of demand for two representative values, $\alpha = 0$ and $\alpha = 0.2$. As $\alpha$ increases, the target population becomes increasingly dominated by high-demand products, reflecting the intended reweighting of the population distribution.

\section{Primal-Forecaster Ablations} \label{app:ablations}
This appendix presents ablations for the primal forecaster $\psi_\theta$. We use these experiments to check whether $\lambda$-based coordination is identifiable from simulated rollouts and to isolate the modeling components needed for accurate aggregate response prediction. Specifically, we evaluate whether the broadcast cost trajectory carries predictive information, and whether population-aware representations and shifted-population training improve robustness under composition shift.

\subsection{Cost-Signal Conditioning}

A necessary condition for $\lambda$-based coordination is that the broadcast cost trajectory has a measurable effect on aggregate utilization. If aggregate inbound were predictable from state features alone, then learning a cost-conditioned response map would provide little value. We therefore compare each cost-conditioned primal forecaster against an unconstrained variant that does not receive $\lambda_{t:t+L}$ as input.

We evaluate both variants on constrained simulator rollouts under two target populations: the baseline population ($\alpha=0$) and a higher-demand population ($\alpha=0.2$). Table~\ref{tab:unconstrained_degradation} reports the MAPE degradation from removing the cost input, relative to the corresponding cost-conditioned forecaster. Larger values indicate that the cost trajectory explains variation in aggregate response that cannot be recovered from state features alone.

Across both populations, removing $\lambda_{t:t+L}$ significantly degrades accuracy, especially during constrained periods. This indicates that the broadcast cost signal induces observable changes in aggregate utilization, supporting the feasibility of learning a cost-conditioned primal response map for capacity coordination.

\begin{table}[htbp]
\centering
\caption{\small Accuracy degradation (absolute percentage-point increase in MAPE) of unconstrained forecast variants under constrained cost trajectories, relative to their cost-conditioned counterparts. Full Backtest Horizon: MAPE over the entire one-year evaluation period. Constrained Period: MAPE restricted to time steps where the broadcast cost $\lambda_t > 0$ (capacity is actively binding). Larger values indicate greater accuracy degradation.}
\label{tab:unconstrained_degradation}
\resizebox{\textwidth}{!}{%
\begin{tabular}{@{}lrrrr@{}}
\toprule
& \multicolumn{2}{c}{\textbf{Target Population 1 ($\alpha=0$)}}
& \multicolumn{2}{c}{\textbf{Target Population 2 ($\alpha=0.2$)}} \\
\cmidrule(lr){2-3} \cmidrule(lr){4-5}
\textbf{Model Class}
& \textbf{Full Backtest} & \textbf{Constrained}
& \textbf{Full Backtest} & \textbf{Constrained} \\
& \textbf{Horizon} & \textbf{Period}
& \textbf{Horizon} & \textbf{Period} \\
\midrule
Bottom-Up & $4.2 \pm 1.1$ & $22.1 \pm 3.05$ & $2.8 \pm 0.9$ & $12.7 \pm 3.0$ \\
Global Aggregate & $1.5 \pm 0.8$ & $37.9 \pm 4.2$ & $1.1 \pm 0.7$ & $8.9 \pm 2.9$ \\
Population-Embedding (per-Agent) Aggregate & $3.7 \pm 0.9$ & $35.2 \pm 3.1$ & $1.2 \pm 0.9$ & $9.1 \pm 2.8$ \\
Population-Embedding (Bucketized) Aggregate & $3.2 \pm 1.0$ & $30.7 \pm 3.6$ & $1.3 \pm 0.9$ & $5.4 \pm 2.7$ \\
\bottomrule
\end{tabular}}
\end{table}

\subsection{Effect of Population-Aware Architecture and Training Coverage}

We next disentangle the effects of population-aware architecture and population-shift training coverage for primal aggregate forecasting. We compare the Global Aggregate model and two Population-Embedding variants, each trained either on the baseline population only or with explicit population-shift coverage. The Bottom-Up model is included as a product-level benchmark. Because it predicts each product independently and aggregates by summation, it naturally applies to any target population whose product-level features are covered by the training distribution. Evaluation is performed on both the baseline population ($\alpha=0$) and a higher-demand population ($\alpha=0.2$), using two fixed capacity trajectories shared across all models.

Table~\ref{tab:arch_train_ablation} reports MAPE for each setting. Models trained only on the baseline population degrade under the shifted population, even when they use population-aware architectures. Conversely, shift coverage improves all aggregate models, but Global Aggregate remains less robust than population-aware variants because it lacks an explicit representation of product composition. The strongest performance is achieved by combining population-aware architecture with population-shift coverage, indicating that both representation and data coverage are necessary for robust generalization.

\begin{table}[htbp]
\centering
\caption{\small Architecture and training ablation (primal interface).
MAPE $(\text{mean} \pm 95\%\,\text{CI})$ under baseline ($\alpha=0$) and
shifted ($\alpha=0.2$) target populations, across two fixed capacity trajectories.
``$+$ Coverage.'' denotes models trained with population-shift coverage
($\alpha \in [-0.5, 0.5]$). Rows above the mid-rule are population-unaware
baselines; rows below are population-aware interfaces.}
\label{tab:arch_train_ablation}
\resizebox{\textwidth}{!}{%
\begin{tabular}{@{}lrrrr@{}}
\toprule
& \multicolumn{2}{c}{\textbf{Target Population 1 ($\alpha=0$)}} & \multicolumn{2}{c}{\textbf{Target Population 2 ($\alpha=0.2$)}} \\
\cmidrule(lr){2-3} \cmidrule(lr){4-5}
\textbf{Model} & \textbf{Scenario 1} & \textbf{Scenario 2} & \textbf{Scenario 1} & \textbf{Scenario 2} \\
\midrule
Global Aggregate & $10.1 \pm 0.9$ & $20.5 \pm 1.1$ & $29.7 \pm 1.4$ & $29.6 \pm 1.4$ \\
Global Aggregate + Coverage & $8.5 \pm 0.7$ & $8.3 \pm 0.7$ & $23.0 \pm 1.2$ & $23.2 \pm 1.2$ \\
Bottom-Up & $9.9 \pm 0.8$ & $8.2 \pm 0.7$ & $9.0 \pm 0.8$ & $8.3 \pm 0.7$ \\
\midrule
Population-Embedding \ (per-Agent) Aggregate & $7.4 \pm 0.7$ & $12.9 \pm 1.0$ & $9.6 \pm 0.8$ & $12.8 \pm 1.0$ \\
Population-Embedding \ (per-Agent) Aggregate\ + Coverage & $6.4 \pm 0.6$ & $9.3 \pm 0.8$ & $7.8 \pm 0.7$ & $9.7 \pm 0.8$ \\
Population-Embedding \ (Bucketized) Aggregate & $8.3 \pm 0.7$ & $11.8 \pm 0.9$ & $7.9 \pm 0.7$ & $10.3 \pm 0.9$ \\
Population-Embedding \ (Bucketized) Aggregate\ + Coverage & $6.2 \pm 0.6$ & $8.3 \pm 0.7$ & $7.0 \pm 0.6$ & $8.7 \pm 0.8$ \\
\bottomrule
\end{tabular}%
}
\end{table}

\section{Additional Robustness and Control Results} \label{app:additional_eval}
This appendix provides additional results supporting the population-shift and capacity-control experiments in Section~\ref{sec:experiments}. The first two subsections evaluate aggregate utilization forecasting under composition shift: we report calibration under demand-based shifts and accuracy under unit-economics shifts. The remaining subsections evaluate capacity-control quality: we report the full set of dual violation metrics across population shifts and compare primal and dual methods under representative low-, baseline-, and high-demand populations.

\subsection*{Utilization Forecast Calibration under Demand-Based Population Shifts}

In addition to accuracy, we also check calibration behavior under demand-based population shifts. We use a regression calibration metric in which the realized aggregate inbound is regressed on the model's forecast; a slope of 1 indicates perfect calibration, values below 1 indicate overcalibration, and values above 1 indicate undercalibration.

\begin{figure}[htbp]
\centering
\includegraphics[width=0.7\textwidth]{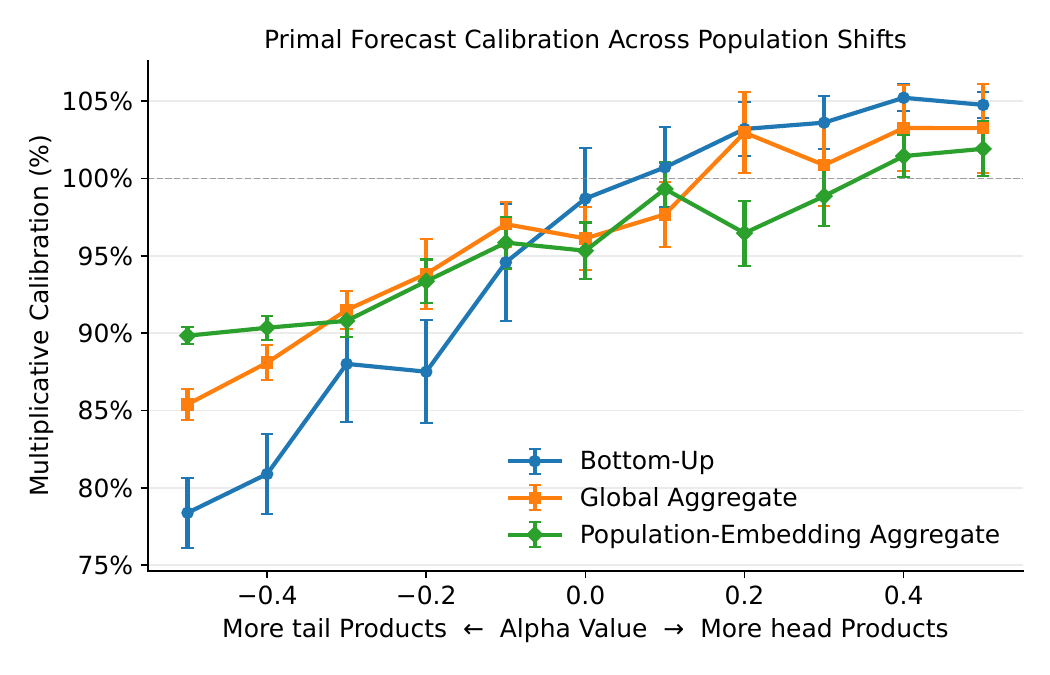}
\caption{\small Calibration slopes of aggregate utilization forecasts under demand-based population
shifts. Slopes closer to 1 indicate better calibration between predicted and realized inbound.}
\label{fig:calibration_shifts}
\end{figure}

Figure~\ref{fig:calibration_shifts} shows that Population-Embedding models maintain slopes closer to 1 across most shifts, typically in the 90--100\% range. In contrast, the Bottom-Up and Global Aggregate models exhibit larger calibration deviations under extreme shifts, consistent with the accuracy degradation observed in Section~\ref{sec:adaptation_experiments}.

\subsection*{Utilization Forecast Accuracy under Unit-Economics Population Shifts}

\begin{figure}[htbp]
\centering
\includegraphics[width=0.7\textwidth]{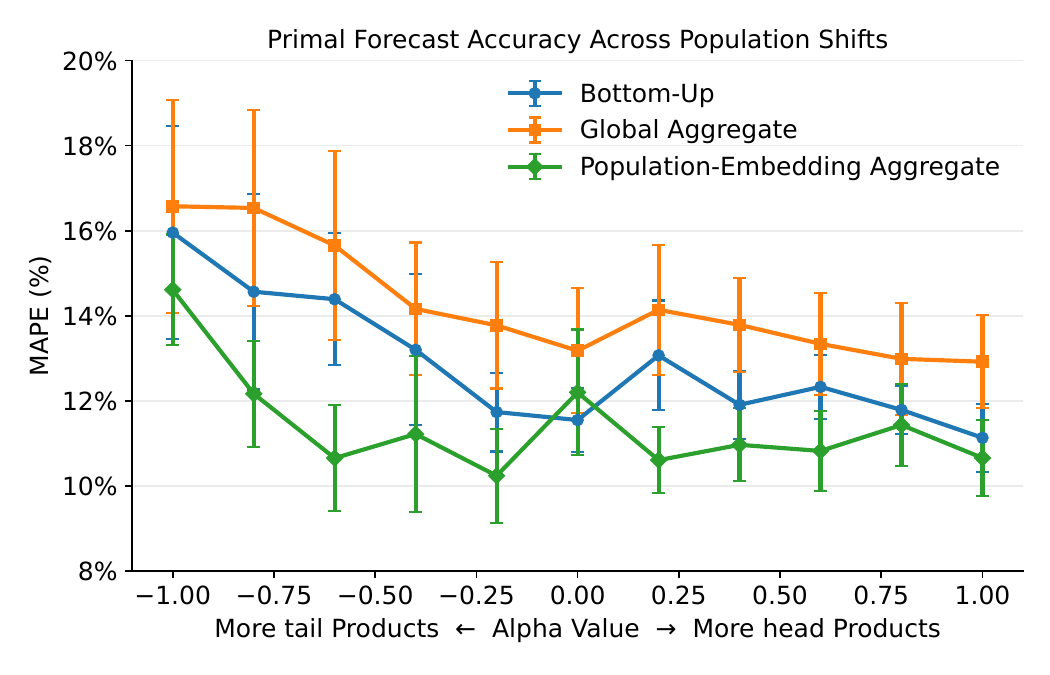}
\caption{\small Aggregate inbound MAPE across unit-economics population shifts. Error bars show 95\% confidence intervals across sampled capacity scenarios.}
\label{fig:unit_economics_shifts}
\end{figure}

Appendix~\ref{app:population_shift} identifies both demand and unit economics as response-relevant attributes. While the main experiments use demand-based shifts, Figure~\ref{fig:unit_economics_shifts} evaluates robustness under shifts along unit economics. The degradation pattern is milder than under demand shifts: all models remain broadly stable, with noticeable degradation mainly in extreme tail-heavy populations. Population-Embedding models achieve the lowest errors across most shift levels, while Global Aggregate remains consistently less accurate.

\subsection*{Dual-Control Violations across Population Shifts}

\begin{figure}[htbp]
\centering
\includegraphics[width=0.9\textwidth]{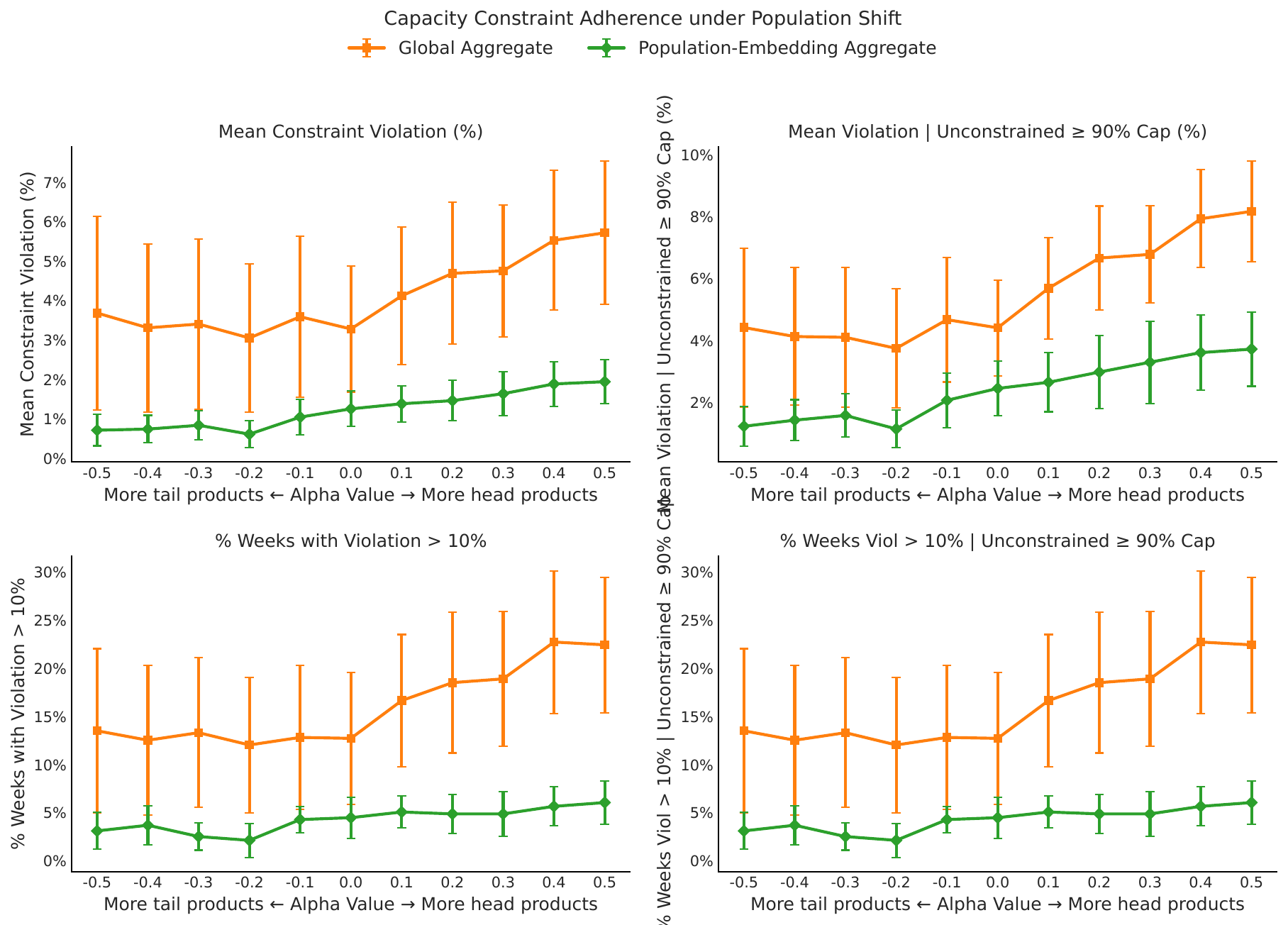}
\caption{\small Violation metrics for the dual coordination interface across
$\alpha$-shifted population distributions. Each point corresponds to one sampled
capacity scenario; lower violation indicates better capacity adherence.}
\label{fig:dual_pop_shift_violation}
\end{figure}

Figure~\ref{fig:dual_pop_shift_violation} reports all four dual control metrics
defined in Section~\ref{sec:experiments}---mean violation, near-limit~(NL) mean
violation, $\mathrm{Viol}_{10}$, and $\mathrm{NL}\text{-}\mathrm{Viol}_{10}$---for
the dual coordination interface evaluated under $\alpha$-shifted populations.
Across all metrics and all shift levels, Population-Embedding interfaces
 consistently achieve lower violations than the Global Aggregate baseline.
These results extend the main-text findings (Section~\ref{sec:control_experiments})
and confirm that population-aware architecture benefits dual control quality across
the full range of population shifts.

\subsection*{Capacity-Control Metrics under Representative Shifts}

Table~\ref{tab:control_alpha_all} reports the full capacity-control metrics for representative low-demand ($\alpha=-0.5$), baseline ($\alpha=0$), and high-demand ($\alpha=0.5$) populations. The trends match Section~\ref{sec:control_experiments}: dual prediction outperforms primal search, and population-aware embeddings improve control quality within both interface families. The gap is largest in the high-demand population, where aggregate response is most sensitive to the cost signal.

\begin{table}[htbp]
\centering
\caption{\small Capacity-control adherence across interface types under representative population shifts. All entries are percentages (lower is better). Mean and NL-Mean report average percentage capacity violation across all weeks and near-limit weeks, respectively. Viol$_{10}$ and NL-Viol$_{10}$ report the percentage of rollout weeks with violation greater than 10\%, over all weeks and near-limit weeks, respectively.}
\label{tab:control_alpha_all}
\small
\begin{tabular}{lccccc}
\toprule
Method & Mean & NL-Mean & Viol$_{10}$ & NL-Viol$_{10}$  \\
\midrule
\multicolumn{5}{c}{Low-demand population ($\alpha = -0.5$, mean demand $\approx 20$)} \\
\midrule
Dual / Global Aggregate  & 0.3 & 0.5 & 0.8  & 1.3  \\
Dual / Embedding         & \textbf{0.2} & \textbf{0.5} & \textbf{0.6} & \textbf{1.5}  \\
Primal / Bottom-Up       & 4.9 & 9.5 & 17.5 & 23.6 \\
Primal / Embedding       & 2.9 & 6.5 & 9.2  & 15.6 \\
\midrule
\multicolumn{5}{c}{Baseline population ($\alpha = 0.0$, mean demand $\approx 86$)} \\
\midrule
Dual / Global Aggregate  & 2.0 & 3.7 & 7.5  & 14.2   \\
Dual / Embedding  & \textbf{1.6} & \textbf{3.1} & \textbf{5.5} & \textbf{10.4}  \\
Primal / Bottom-Up    & 8.6 & 13.4 & 22.0 & 34.1  \\
Primal / Embedding & 4.2 & 7.7  & 11.0 & 18.7  \\
\midrule
\multicolumn{5}{c}{High-demand population ($\alpha = 0.5$, mean demand $\approx 635$)} \\
\midrule
Dual / Global Aggregate  & 8.0 & 11.0 & 32.0 & 41.0   \\
Dual / Embedding         & \textbf{4.0} & \textbf{6.0} & \textbf{17.0} & \textbf{25.5} \\
Primal / Bottom-Up       & 22.0 & 28.0 & 50.0 & 61.1 \\
Primal / Embedding       & 7.0  & 10.0 & 23.0 & 30.0  \\
\bottomrule
\end{tabular}
\end{table}

\section{Sim2Real Evaluation of Simulation-Based Primal Forecasting} \label{app:sim2real_ablation}
This appendix evaluates whether simulator-generated rollouts are a useful training source for primal response forecasters. We organize the analysis around five questions. First, does the simulator reproduce real aggregate inbound dynamics closely enough to serve as a data source? Second, do simulator rollouts provide a stable supervised learning signal? Third, do simulator-trained forecasters transfer to real aggregate inbound observations under matched evaluation settings? Fourth, can lightweight calibration with early real data reduce residual Sim2Real bias? Finally, as a secondary use case, can simulation backtests preserve the relative ranking of forecaster classes before deployment?

\subsection{Evaluation Setup} \label{app:sim2real_setup}

We evaluate Sim2Real transfer in the same large-scale supply-chain capacity-control setting studied in the main text. The evaluation uses two real-world operational studies, denoted \textbf{Study~1} and \textbf{Study~2}. Both studies come from deployments at a major retailer, were conducted in different years, and involve hundreds of thousands of active product agents. They correspond to different real evaluation periods and deployed policy regimes, but the forecasting task is identical in both cases: given state, population, and cost trajectories, predict aggregate inbound utilization.

For each study, the simulator is initialized from the corresponding real system state and rolled out under matched exogenous inputs and coordination signals. We use these matched simulator and real trajectories for several evaluations. Sections~\ref{app:sim2real_fidelity}--\ref{app:sim2real_learnability} evaluate simulator fidelity and the learnability of simulator-generated trajectories. Sections~\ref{app:sim2real_transfer} and~\ref{app:sim2real_adaptation} evaluate cold-start transfer to real observations and lightweight calibration with early real data.

These first four evaluations use two baseline forecaster classes from Section~\ref{sec:model_class}: Global Aggregate and Bottom-Up. This is sufficient for validating the simulator as a training source because the goal is to test whether simulated trajectories are coherent and transferable at both aggregate and product-summed forecasting granularities, rather than to compare all architecture variants. Global Aggregate predicts total inbound directly from population-level features, while Bottom-Up predicts inbound at the individual-product level and aggregates by summation.

Finally, Section~\ref{app:sim2real_backtest} evaluates simulation as a model-selection backtest by comparing the relative performance of all forecaster classes, including the Population-Embedding variants from Section~\ref{sec:model_class}.

\paragraph{Forecast horizons.}
All models produce multi-horizon forecasts; accuracy is reported across four forecast horizon groups: \textbf{LT=1} (one week ahead), \textbf{LT=2}, \textbf{LT=3}, and \textbf{LT=4--10} (four to ten weeks ahead).

\paragraph{Reference baseline.}
As a reference, we compare against an existing simulation-based inbound forecast evaluated on similar cohorts within the same real-world study periods (Table~\ref{tab:ipc_sim_baseline}). This reference is not a direct comparison because it comes from a separate simulation-based forecasting pipeline, but it provides an operational benchmark for the same broad forecasting task.

\begin{table}[htbp]
\centering
\caption{\small Accuracy of the simulation-based reference forecast on evaluation cohorts from the corresponding real-world study periods. Reported values are MAPE (mean $\pm$ 95\% CI) across forecast horizons.}
\label{tab:ipc_sim_baseline}
\begin{tabular}{lrrrr}
\toprule
\textbf{Setting} & \textbf{LT=1} & \textbf{LT=2} & \textbf{LT=3} & \textbf{LT=4--10} \\
\midrule
Study~1 & $15.9\pm 3.8$ & $14.7\pm 3.9$ & $16.7\pm 2.8$ & $22.1\pm 3.0$ \\
Study~2 & $14.3 \pm 14.9$ & $21.5 \pm 20.0$ & $27.7 \pm 21.3$ & $40.6 \pm 12.6$ \\
\bottomrule
\end{tabular}
\end{table}

\subsection{Simulator Fidelity} \label{app:sim2real_fidelity}

We first assess whether the simulator reproduces realistic aggregate inbound dynamics by comparing simulated inbound with realized inbound when initialized from real system states. This analysis does not involve any forecasting model; it evaluates only the fidelity of the simulated trajectories.

In Study~1, direct simulator rollouts achieve aggregate inbound MAPE of $13.24 \pm 2.43\%$. In Study~2, simulator rollouts achieve $11.79 \pm 2.62\%$. In both studies, simulator rollouts are comparable to or better than the simulation-based reference forecast, indicating that the simulator captures aggregate response dynamics with reasonable fidelity.

\subsection{Learnability of Simulated Dynamics} \label{app:sim2real_learnability}

We next evaluate whether simulator-generated trajectories provide a coherent and learnable training signal for supervised inbound forecasting. Models are trained on simulator rollouts and evaluated on temporally held-out simulator data from the same study.

As a reference, we also evaluate models trained on real data, using the same model classes and training procedures. This comparison is not a Sim2Real transfer evaluation; rather, it checks whether simulator-generated trajectories are at least as stable and learnable as standard supervised forecasting data.

As shown in Table~\ref{tab:learnability}, simulator-trained models achieve stable errors on held-out simulator trajectories. The error ranges are broadly comparable to those of models trained and evaluated on historical real data, suggesting that simulator-generated trajectories provide an internally consistent and learnable supervised training distribution.

\begin{table}[htbp]
\centering
\caption{\small Learnability of simulator-generated dynamics. Models are trained on simulator rollouts and evaluated on held-out simulator data from the same study. Historical real-data baselines are trained on pre-study historical observations and evaluated on held-out real data from the Study~2 evaluation period, using the same model classes and training procedures. Reported values are MAPE (mean $\pm$ 95\% CI).}
\label{tab:learnability}
\resizebox{\textwidth}{!}{
\begin{tabular}{lllrrrr}
\toprule
\textbf{Training Source} & \textbf{Eval Source} & \textbf{Model} & \textbf{LT=1} & \textbf{LT=2} & \textbf{LT=3} & \textbf{LT=4--10} \\
\midrule
Simulator (Study~1) & Simulator (Study~1) & Global Aggregate & $9.2 \pm 1.8$ & $9.2 \pm 1.6$ & $10.8 \pm 2.0$ & $11.1 \pm 2.0$ \\
Simulator (Study~1) & Simulator (Study~1) & Bottom-Up & $12.2 \pm 3.1$ & $11.5 \pm 2.7$ & $10.7 \pm 2.5$ & $9.6 \pm 0.9$ \\
\midrule
Simulator (Study~2) & Simulator (Study~2) & Global Aggregate & $10.0 \pm 1.4$ & $10.5 \pm 1.5$ & $11.9 \pm 1.6$ & $9.1 \pm 1.3$ \\
Simulator (Study~2) & Simulator (Study~2) & Bottom-Up & $10.0 \pm 2.6$ & $12.9 \pm 2.8$ & $14.0 \pm 3.0$ & $12.2 \pm 1.7$ \\
\midrule
Historical Real Data & Held-out Real Data & Global Aggregate & $11.7 \pm 1.5$ & $11.9 \pm 1.5$ & $11.8 \pm 1.4$ & $11.5 \pm 1.3$ \\
Historical Real Data & Held-out Real Data & Bottom-Up & $8.9 \pm 2.2$ & $9.9 \pm 2.3$ & $10.5 \pm 2.4$ & $9.7 \pm 1.2$ \\
\bottomrule
\end{tabular}}
\end{table}

\subsection{Sim2Real Generalization} \label{app:sim2real_transfer}

We test whether models trained on simulated data generalize to real aggregate inbound observations. For each study, models are trained on simulator-generated trajectories from the pre-evaluation period and tested on realized inbound during the matched real-world evaluation window. These simulator rollouts are on-policy for the study setting: the simulator runs the same policy evaluated in the real study, starting from the corresponding real system state and using matched exogenous inputs and coordination signals.

As a comparison baseline, we train models on historical real data collected under prior operating policies and evaluate them on the same real-world cohorts. This baseline is off-policy relative to the study policy, and helps quantify the importance of on-policy training data for learning the target response dynamics.

Across both studies, simulator-trained models achieve lower MAPE on real outcomes than historical real-data baselines in most lead-time groups (Table~\ref{tab:sim2real_transfer}). These results highlight the importance of on-policy response data for Sim2Real forecasting: although historical data are real observations, they may not capture the aggregate inbound dynamics induced by the launched policy.

\begin{table}[htbp]
\centering
\caption{\small Sim2Real generalization. Models are trained on simulator rollouts and evaluated on realized inbound during the matched real-world evaluation periods. Historical real-data baselines are trained on past real observations and evaluated on the same real-world cohorts. Reported values are MAPE (mean $\pm$ 95\% CI) by lead-time group.}
\label{tab:sim2real_transfer}
\resizebox{\textwidth}{!}{
\begin{tabular}{lllrrrr}
\toprule
\textbf{Training Source} & \textbf{Eval Source} & \textbf{Model} & \textbf{LT=1} & \textbf{LT=2} & \textbf{LT=3} & \textbf{LT=4--10} \\
\midrule
Simulator (Study~1) & Real (Study~1) & Global Aggregate & $13.3\pm 2.7$ & $11.7\pm 2.8$ & $13.2\pm 3.1$ & $14.8\pm 2.8$ \\
Simulator (Study~1) & Real (Study~1) & Bottom-Up & $13.4\pm 3.0$ & $13.3\pm 3.0$ & $13.4\pm 2.8$ & $13.4\pm 1.1$ \\
\midrule
Simulator (Study~2) & Real (Study~2) & Global Aggregate & $12.7\pm 3.6$ & $15.3\pm 7.4$ & $9.4\pm 3.8$ & $11.5\pm 3.4$ \\
Simulator (Study~2) & Real (Study~2) & Bottom-Up & $16.4\pm 6.2$ & $17.7\pm 7.4$ & $20.1\pm 8.0$ & $19.4\pm 3.9$ \\
\midrule
Off-policy Real Data & Real (Study~1) & Global Aggregate & $15.0\pm 2.9$ & $14.8\pm 3.1$ & $15.2\pm 3.1$ & $16.0\pm 3.1$ \\
Off-policy Real Data & Real (Study~1) & Bottom-Up & $16.7\pm 2.9$ & $16.8\pm 3.0$ & $17.7\pm 3.0$ & $19.5\pm 1.2$ \\
Off-policy Real Data & Real (Study~2) & Global Aggregate & $25.6 \pm 5.2$ & $27.8 \pm 6.3$ & $28.1 \pm 6.4$ & $25.2 \pm 5.7$ \\
\bottomrule
\end{tabular}}
\end{table}

\subsection{Sim2Real Transfer with Real-Data Adaptation} \label{app:sim2real_adaptation}

We apply lightweight calibration using observations collected early in the real-world evaluation period. For each model, we fit a simple regression adjustment from model forecasts to realized inbound outcomes, and then evaluate the calibrated forecasts on the remaining evaluation window.

Table~\ref{tab:sim2real_adaptation} reports forecasting accuracy after calibration for the same models shown in Table~\ref{tab:sim2real_transfer}. Calibration improves several simulator-trained models, especially the Global Aggregate model in Study~2 and the Bottom-Up model at longer lead times. However, gains are not uniform across models or studies. Historical real-data baselines show limited benefit from the same calibration procedure, suggesting that simple post-hoc correction cannot fully compensate for a training distribution that is off-policy relative to the launched policy.

\begin{table}[htbp]
\centering
\caption{\small Forecasting accuracy after regression-based calibration using early real observations. Simulator-trained models are corrected using data from the beginning of the evaluation period. Off-policy historical baselines use the same calibration procedure. Reported values are MAPE (mean $\pm$ 95\% CI).}
\label{tab:sim2real_adaptation}
\resizebox{\textwidth}{!}{
\begin{tabular}{lllrrrr}
\toprule
\textbf{Training Source} & \textbf{Eval Source} & \textbf{Model} & \textbf{LT=1} & \textbf{LT=2} & \textbf{LT=3} & \textbf{LT=4--10} \\
\midrule
Simulator (Study~1) & Real (Study~1) & Global Aggregate + Calib. & $11.8\pm 2.6$ & $11.2\pm 3.1$ & $15.1\pm 3.6$ & $18.8\pm 3.2$ \\
Simulator (Study~1) & Real (Study~1) & Bottom-Up + Calib. & $13.5 \pm 6.0$ & $12.4 \pm 5.7$ & $12.3 \pm 5.4$ & $11.2 \pm 1.8$ \\
\midrule
Simulator (Study~2) & Real (Study~2) & Global Aggregate + Calib. & $12.6 \pm 3.6$ & $13.0 \pm 7.1$ & $6.9 \pm 3.1$ & $9.7 \pm 2.9$ \\
Simulator (Study~2) & Real (Study~2) & Bottom-Up + Calib. & $19.6 \pm 6.8$ & $21.3 \pm 6.3$ & $18.2 \pm 9.1$ & $14.4 \pm 4.7$ \\
\midrule
Off-policy Real Data & Real (Study~1) & Global Aggregate + Calib. & $15.5\pm 3.0$ & $15.9\pm 3.3$ & $15.8\pm 3.6$ & $18.0\pm 3.4$ \\
Off-policy Real Data & Real (Study~1) & Bottom-Up + Calib. & $18.4\pm 3.0$ & $15.4\pm 2.6$ & $15.2\pm 2.3$ & $14.1\pm 1.8$ \\
Off-policy Real Data & Real (Study~2) & Bottom-Up + Calib. & $29.9 \pm 9.6$ & $29.6 \pm 11.7$ & $28.4 \pm 12.9$ & $20.5 \pm 5.5$ \\
\bottomrule
\end{tabular}}
\end{table}

\subsection{Simulation as a Backtesting Tool} \label{app:sim2real_backtest}

Another use of the simulator is pre-deployment backtesting of forecaster classes. Even when absolute simulation error differs from real-world error, the simulator can still be useful if it preserves the relative ranking of models. We evaluate this in Study~2, which corresponds to the cold-start Sim2Real evaluation in Section~\ref{subsec:eval_sim2real}. Unlike the preceding evaluations, this backtest includes all forecaster classes: Bottom-Up, Global Aggregate, Population-Embedding (per-Agent) Aggregate, and Population-Embedding (Bucketized) Aggregate.

Table~\ref{tab:gym_backtest} reports MAPE by lead-time group on simulated trajectories from the matched Study~2 evaluation cohort. The simulation backtest recovers the same qualitative pattern observed in the real-data Sim2Real results (Table~\ref{tab:sim2real_nvf}): Population-Embedding forecasters outperform the Bottom-Up and Global Aggregate baselines. Within the Population-Embedding family, the bucketized variant performs best in simulation. Absolute MAPE differs between simulation and real observations, but the preserved ranking suggests that simulation can provide a useful pre-deployment backtest for selecting forecaster classes before real evaluation data are available.

\begin{table}[htbp]
\centering
\caption{\small Simulation backtest of forecasting models in Study~2, the same setting used for the main-text cold-start Sim2Real evaluation. Models are evaluated on simulated trajectories for the matched evaluation cohort. Reported values are MAPE (mean $\pm$ 95\% CI) across forecast horizons.}
\label{tab:gym_backtest}
\resizebox{\textwidth}{!}{
\begin{tabular}{lrrrr}
\toprule
\textbf{Model} & \textbf{LT=1} & \textbf{LT=2} & \textbf{LT=3} & \textbf{LT=4--10} \\
\midrule
Bottom-Up & $8.4 \pm 2.0$ & $8.0 \pm 1.8$ & $8.8 \pm 2.1$ & $6.6 \pm 0.7$ \\
Global Aggregate & $8.6 \pm 2.3$ & $10.6 \pm 2.3$ & $10.7 \pm 2.6$ & $10.0 \pm 1.3$ \\
Population-Embedding (per-Agent) Aggregate & $7.8 \pm 1.8$ & $8.3 \pm 1.8$ & $8.3 \pm 1.9$ & $7.4 \pm 0.7$ \\
Population-Embedding (Bucketized) Aggregate & $7.3 \pm 1.6$ & $7.0 \pm 1.7$ & $7.1 \pm 1.9$ & $6.3 \pm 0.7$ \\
\bottomrule
\end{tabular}}
\end{table}

\paragraph{Summary.}
Taken together, these results support the use of simulator-generated data for training primal response forecasters that transfer to real aggregate inbound observations. The simulator provides reasonable aggregate fidelity, generates learnable supervised trajectories, and simulator-trained models generally outperform off-policy historical baselines on real observations. Lightweight calibration with early real data can further reduce residual Sim2Real bias in several settings. As a secondary use case, simulation backtests also preserve the qualitative ranking of forecaster classes, supporting their use for pre-deployment model selection. These results are empirical and depend on the calibrated simulator used in this study. Improving simulator calibration and better characterizing the Sim2Real gap are important open directions; we leave systematic approaches to both for future work.

\end{document}